\documentclass[aps, prd,  superscriptaddress, showpacs, longbibliography, floatfix, nofootinbib]{revtex4-2}

\usepackage{bm, amsmath, amssymb, relsize, amsfonts, mathrsfs, multirow, braket, siunitx, color, booktabs}
\usepackage{graphicx}
\graphicspath{{./figures/}}

\DeclareSIUnit \parsec {pc}

\usepackage[dvipsnames]{xcolor}
\usepackage[unicode]{hyperref}
\hypersetup{
  colorlinks=true,
  citecolor=RoyalBlue,
  linkcolor=blue,
  urlcolor=blue
}

\usepackage{array}
\usepackage{makecell}
\usepackage{amsmath}
\usepackage{cancel}
\usepackage{tabularx}

\usepackage[caption=false]{subfig}
\usepackage{array}
\usepackage{booktabs}
\usepackage{tabularx}
\usepackage{booktabs} 
\usepackage{amsmath}  
\usepackage{tabularx}
\usepackage{booktabs}
\usepackage{amsmath}
\usepackage{cancel}
\usepackage{array}
\usepackage{dashrule}
\usepackage{multirow}
\usepackage{arydshln}
\usepackage{cancel}
\usepackage{orcidlink}
\usepackage{graphicx}
\usepackage{slashed}
\usepackage{dcolumn}
\usepackage{bm}
\usepackage{tikz}
\usepackage{hyperref}
\usepackage{paralist}
\usepackage{epsfig}
\usepackage{placeins}
\usepackage{mathrsfs}
\usepackage{comment}
\usepackage{color, colortbl}
\usepackage[inline]{enumitem}
\usepackage{soul}
\usepackage{adjustbox}
\usepackage{amssymb}
\usepackage{caption}
\usepackage{subcaption}
\usepackage{float}
\usepackage{cancel}
\usepackage{verbatim}
\usepackage{amsmath}
\usepackage[toc,page]{appendix}
\usepackage{graphicx}

\DeclareMathAlphabet{\mathpzc}{OT1}{pzc}{m}{it}
	
\definecolor{LightCyan}{rgb}{0.88,1,1}
\definecolor{lightgray}{gray}{0.9}

\def \IITGn     {Department of Physics, Indian Institute of Technology Gandhinagar, Gujarat 382055, India.\vspace*{4pt}}

\begin{document}

\title{Charged Higgs Signatures at Future Electron-Proton Colliders}

\author{\textsc{Baradhwaj Coleppa}\orcidlink{0000-0002-8761-3138}}
\email{baradhwaj@iitgn.ac.in }
\affiliation{\IITGn}

\author{\textsc{Gokul B. Krishna}\orcidlink{0000-0001-7083-4308}}
\email{gokulb@iitgn.ac.in}
\affiliation{\IITGn}

\begin{abstract}
In this work, we present a detailed collider phenomenology study of the charged Higgs boson within a Beyond the Standard Model (BSM) framework featuring an extended gauge and scalar sector. The charged Higgs can decay via conventional modes, such as \( H^- \to \bar{t}b \) and \( H^- \to W^-h \), as well as through exotic channels like \( H^- \to W'Z \) (or \( WZ' \)). These decays lead to distinct final-state topologies determined by the nature of the intermediate particles. We perform a comprehensive phenomenological analysis at future electron-proton colliders, namely the LHeC and FCC-eh, considering the luminosity projections provided in their design reports. Our results indicate that the conventional decay modes of the charged Higgs boson can achieve observable sensitivity and even discovery prospects at sufficiently high luminosities. In contrast, the exotic decay channel \( H^- \to W'Z \) does not exhibit any viable discovery potential. These findings highlight the complementarity of future electron-proton colliders in probing extended Higgs sectors, particularly through conventional charged Higgs signatures.
\end{abstract}

\maketitle

\section{Introduction}

The pursuit of physics beyond the Standard Model (SM) remains a central focus in contemporary high-energy physics. Extending the SM and testing new physics scenarios through various BSM frameworks at current and future colliders continues to be a promising path toward uncovering novel phenomena. The Large Hadron Collider (LHC) has served as a powerful testing ground for such theories, with its crowning achievement being the discovery of the Higgs boson by the ATLAS~\cite{ATLAS:2012yve} and CMS~\cite{CMS:2012qbp} collaborations completing the particle content of the SM. In the ongoing search for new physics, it is both natural and necessary to explore BSM signatures in alternative collider environments. Compared to the complex hadronic setting of proton--proton collisions at the LHC, future electron--proton (\(e^-p\)) colliders, such as the proposed Large Hadron–Electron Collider (LHeC)~\cite{LHeCStudyGroup:2012zhm} and Future Circular Collider in electron–hadron mode (FCC-eh)~\cite{Adolphsen:2022ibf}, offer a significantly cleaner experimental environment with a well-defined initial state. The suppression of QCD backgrounds, absence of pile-up, and simpler final-state topologies in \(e^-p\) collisions allow for more precise kinematic reconstructions and enhanced sensitivity to new physics, particularly in scenarios involving lepton--quark interactions. These advantages make \(e^-p\) colliders an attractive and complementary avenue for probing BSM physics.

Among various BSM scenarios, an especially compelling direction—distinct from the more conventional scalar-sector extensions such as the Two-Higgs Doublet Models (2HDM)~\cite{Branco:2011iw}—involves extending the SM gauge group in together with introducing additional scalar fields. Such frameworks naturally lead to richer patterns of symmetry breaking and a more diverse phenomenology. Models of this type have been explored in works like~\cite{Coleppa:2020set}, where the charged Higgs sector exhibits significant modifications. Compared to the 2HDM, these models can feature new interaction vertices that are entirely absent in simpler setups, thereby allowing exotic decay patterns of the form \textit{New Physics (NP)} $\to$ \textit{NP}.\footnote{In particular, the charged Higgs $H^\pm$ can decay into other exotic states that themselves carry NP content—an option unavailable in minimal scenarios like the 2HDM.} As a result, such models predict distinctive decay channels for the charged Higgs boson that go beyond those typically expected in minimal extensions. Recent studies~\cite{Coleppa:2021wjx} have further highlighted that these exotic signatures could potentially be probed at the LHC, provided sufficiently high luminosity.

Motivated by these findings, we explore the discovery prospects of the charged Higgs boson in extended gauge models at future \(e^-p\) colliders. Facilities like the LHeC, with proposed center-of-mass energy \(\sqrt{s} = 1.3~\text{TeV}\), and the FCC-eh, with \(\sqrt{s} = 3.5~\text{TeV}\), offer unique opportunities to study such signals in a cleaner environment, potentially enhancing the discovery sensitivity for BSM Higgs sectors. The structure of this paper is as follows: in \autoref{tab:model}, we present the essential setup for our study, including collider parameters and theoretical considerations. In \autoref{tab:pheno}, we detail the phenomenological analysis, including signal versus background comparisons for various viable decay modes of the charged Higgs boson. Up until this point, we strive to keep our discussions model-independent, assuming only a spectrum that includes additional gauge and scalar degrees of freedom beyond the SM. To interpret these findings in the context of a concrete model, we then comment on how the results apply to the extended gauge model outlined in Ref.~\cite{Coleppa:2020set}, as summarized in \autoref{tab:PI}. Finally, in \autoref{tab:conclusion}, we conclude with a summary of our main findings.

\section{The $H^\pm$ Boson Discovery Prospects at Future Colliders}
\label{tab:model}

\subsection{Set-up and Current Limits}

As stated in the introduction, we adopt a model-independent approach in this study, making only a minimal set of assumptions inspired by the framework presented in Ref.~\cite{Coleppa:2021wjx}. In particular, we consider the existence of a heavy charged Higgs bosons \( H^\pm \), in addition to the neutral SM-like Higgs boson \( h^0 \). A distinctive feature of the scenario under consideration is the presence of new heavy neutral and charged gauge bosons, denoted as \( Z' \) and \( W' \), which are assumed to be fermiophobic. This latter assumption is crucial as it helps to evade direct collider limits on these heavy gauge bosons. To enable a meaningful exploration of the associated decay channels, we assume that the masses of these heavy gauge bosons lie within a range that remains consistent with current experimental constraints, while still allowing for \( H^\pm \) decays into these new states. Of particular interest are decay topologies involving the couplings \( H^\pm W^{'\mp} Z \) and \( H^\pm W^{\mp} Z' \), which yield signatures beyond those of more conventional models. Although such exotic decay modes are theoretically intriguing, our detailed analysis shows that they do not offer any realistic discovery prospects at future \( e^-p \) colliders such as the LHeC and FCC-eh, even with their projected luminosities. For this reason, we do not include this decay mode in the main phenomenological analysis. However, for completeness, we briefly discuss its features and explain why it lacks discovery potential in \autoref{app} of the Appendix.

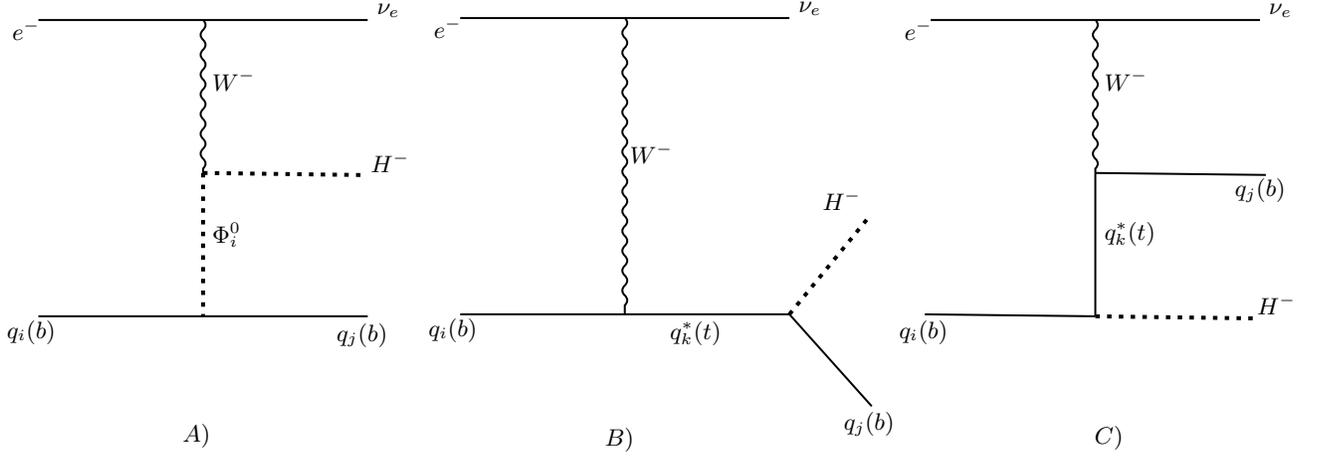
\begin{figure}
    \centering
\tikzset{every picture/.style={line width=0.75pt}} 

\begin{tikzpicture}[x=0.75pt,y=0.75pt,yscale=-1,xscale=1]

\draw    (17.74,114.56) -- (183.81,114.56) ;
\draw    (100.78,114.56) .. controls (102.45,116.23) and (102.45,117.89) .. (100.78,119.56) .. controls (99.11,121.23) and (99.11,122.89) .. (100.78,124.56) .. controls (102.45,126.23) and (102.45,127.89) .. (100.78,129.56) .. controls (99.11,131.23) and (99.11,132.89) .. (100.78,134.56) .. controls (102.45,136.23) and (102.45,137.89) .. (100.78,139.56) .. controls (99.11,141.23) and (99.11,142.89) .. (100.78,144.56) .. controls (102.45,146.23) and (102.45,147.89) .. (100.78,149.56) .. controls (99.11,151.23) and (99.11,152.89) .. (100.78,154.56) .. controls (102.45,156.23) and (102.45,157.89) .. (100.78,159.56) .. controls (99.11,161.23) and (99.11,162.89) .. (100.78,164.56) .. controls (102.45,166.23) and (102.45,167.89) .. (100.78,169.56) .. controls (99.11,171.23) and (99.11,172.89) .. (100.78,174.56) .. controls (102.45,176.23) and (102.45,177.89) .. (100.78,179.56) .. controls (99.11,181.23) and (99.11,182.89) .. (100.78,184.56) .. controls (102.45,186.23) and (102.45,187.89) .. (100.78,189.56) -- (100.78,191.77) -- (100.78,191.77) ;
\draw [line width=1.5]  [dash pattern={on 1.69pt off 2.76pt}]  (100.78,191.77) -- (100.78,264.12) ;
\draw [line width=1.5]  [dash pattern={on 1.69pt off 2.76pt}]  (100.78,191.77) -- (183.24,192.85) ;
\draw    (17.74,264.12) -- (183.81,264.12) ;
\draw    (467.74,114.56) -- (633.81,114.56) ;
\draw    (550.78,114.56) .. controls (552.45,116.23) and (552.45,117.89) .. (550.78,119.56) .. controls (549.11,121.23) and (549.11,122.89) .. (550.78,124.56) .. controls (552.45,126.23) and (552.45,127.89) .. (550.78,129.56) .. controls (549.11,131.23) and (549.11,132.89) .. (550.78,134.56) .. controls (552.45,136.23) and (552.45,137.89) .. (550.78,139.56) .. controls (549.11,141.23) and (549.11,142.89) .. (550.78,144.56) .. controls (552.45,146.23) and (552.45,147.89) .. (550.78,149.56) .. controls (549.11,151.23) and (549.11,152.89) .. (550.78,154.56) .. controls (552.45,156.23) and (552.45,157.89) .. (550.78,159.56) .. controls (549.11,161.23) and (549.11,162.89) .. (550.78,164.56) .. controls (552.45,166.23) and (552.45,167.89) .. (550.78,169.56) .. controls (549.11,171.23) and (549.11,172.89) .. (550.78,174.56) .. controls (552.45,176.23) and (552.45,177.89) .. (550.78,179.56) .. controls (549.11,181.23) and (549.11,182.89) .. (550.78,184.56) .. controls (552.45,186.23) and (552.45,187.89) .. (550.78,189.56) -- (550.78,191.77) -- (550.78,191.77) ;
\draw [line width=1.5]  [dash pattern={on 1.69pt off 2.76pt}]  (550.78,264.12) -- (573.99,264.43) -- (633.24,265.2) ;
\draw    (464.78,263.12) -- (550.78,264.12) ;
\draw    (550.78,191.77) -- (636.78,192.77) ;
\draw    (230.44,113.56) -- (396.5,113.56) ;
\draw    (313.47,113.56) .. controls (315.14,115.23) and (315.14,116.89) .. (313.47,118.56) .. controls (311.8,120.23) and (311.8,121.89) .. (313.47,123.56) .. controls (315.14,125.23) and (315.14,126.89) .. (313.47,128.56) .. controls (311.8,130.23) and (311.8,131.89) .. (313.47,133.56) .. controls (315.14,135.23) and (315.14,136.89) .. (313.47,138.56) .. controls (311.8,140.23) and (311.8,141.89) .. (313.47,143.56) .. controls (315.14,145.23) and (315.14,146.89) .. (313.47,148.56) .. controls (311.8,150.23) and (311.8,151.89) .. (313.47,153.56) .. controls (315.14,155.23) and (315.14,156.89) .. (313.47,158.56) .. controls (311.8,160.23) and (311.8,161.89) .. (313.47,163.56) .. controls (315.14,165.23) and (315.14,166.89) .. (313.47,168.56) .. controls (311.8,170.23) and (311.8,171.89) .. (313.47,173.56) .. controls (315.14,175.23) and (315.14,176.89) .. (313.47,178.56) .. controls (311.8,180.23) and (311.8,181.89) .. (313.47,183.56) .. controls (315.14,185.23) and (315.14,186.89) .. (313.47,188.56) .. controls (311.8,190.23) and (311.8,191.89) .. (313.47,193.56) .. controls (315.14,195.23) and (315.14,196.89) .. (313.47,198.56) .. controls (311.8,200.23) and (311.8,201.89) .. (313.47,203.56) .. controls (315.14,205.23) and (315.14,206.89) .. (313.47,208.56) .. controls (311.8,210.23) and (311.8,211.89) .. (313.47,213.56) .. controls (315.14,215.23) and (315.14,216.89) .. (313.47,218.56) .. controls (311.8,220.23) and (311.8,221.89) .. (313.47,223.56) .. controls (315.14,225.23) and (315.14,226.89) .. (313.47,228.56) .. controls (311.8,230.23) and (311.8,231.89) .. (313.47,233.56) .. controls (315.14,235.23) and (315.14,236.89) .. (313.47,238.56) .. controls (311.8,240.23) and (311.8,241.89) .. (313.47,243.56) .. controls (315.14,245.23) and (315.14,246.89) .. (313.47,248.56) .. controls (311.8,250.23) and (311.8,251.89) .. (313.47,253.56) .. controls (315.14,255.23) and (315.14,256.89) .. (313.47,258.56) -- (313.47,263.12) -- (313.47,263.12) ;
\draw [line width=1.5]  [dash pattern={on 1.69pt off 2.76pt}]  (396.5,263.12) -- (437,213.5) ;
\draw    (230.44,263.12) -- (396.5,263.12) ;
\draw    (396.5,263.12) -- (438,309.5) ;
\draw    (550.78,191.77) -- (550.78,264.12) ;

\draw (103.91,138.32) node [anchor=north west][inner sep=0.75pt]    {$W^{-}$};
\draw (103.87,215.11) node [anchor=north west][inner sep=0.75pt]    {$\Phi _{i}^{0}$};
\draw (3.02,112.88) node [anchor=north west][inner sep=0.75pt]    {$e^{-}$};
\draw (0.31,264.62) node [anchor=north west][inner sep=0.75pt]    {$q_{i}( b)$};
\draw (183.97,179.69) node [anchor=north west][inner sep=0.75pt]    {$H^{-}$};
\draw (166.69,266.08) node [anchor=north west][inner sep=0.75pt]    {$q_{j}( b)$};
\draw (187,105) node [anchor=north west][inner sep=0.75pt]    {$\nu _{e}$};
\draw (432,154) node [anchor=north west][inner sep=0.75pt]   [align=left] { };
\draw (553.91,138.32) node [anchor=north west][inner sep=0.75pt]    {$W^{-}$};
\draw (553.87,215.11) node [anchor=north west][inner sep=0.75pt]    {$q_{k}^{*}( t)$};
\draw (453.02,112.88) node [anchor=north west][inner sep=0.75pt]    {$e^{-}$};
\draw (450.31,264.62) node [anchor=north west][inner sep=0.75pt]    {$q_{i}( b)$};
\draw (630.97,251.69) node [anchor=north west][inner sep=0.75pt]    {$H^{-}$};
\draw (619.69,193.08) node [anchor=north west][inner sep=0.75pt]    {$q_{j}( b)$};
\draw (637,105) node [anchor=north west][inner sep=0.75pt]    {$\nu _{e}$};
\draw (314.6,175.32) node [anchor=north west][inner sep=0.75pt]    {$W^{-}$};
\draw (215.71,111.88) node [anchor=north west][inner sep=0.75pt]    {$e^{-}$};
\draw (213,263.62) node [anchor=north west][inner sep=0.75pt]    {$q_{i}( b)$};
\draw (411.67,198.69) node [anchor=north west][inner sep=0.75pt]    {$H^{-}$};
\draw (422.38,311.08) node [anchor=north west][inner sep=0.75pt]    {$q_{j}( b)$};
\draw (399.69,104) node [anchor=north west][inner sep=0.75pt]    {$\nu _{e}$};
\draw (334.87,265.11) node [anchor=north west][inner sep=0.75pt]    {$q_{k}^{*}( t)$};
\draw (89,317) node [anchor=north west][inner sep=0.75pt]    {$A)$};
\draw (549,318) node [anchor=north west][inner sep=0.75pt]    {$C)$};
\draw (302,319) node [anchor=north west][inner sep=0.75pt]    {$B)$};

\end{tikzpicture}
\caption{Representative Feynman diagrams for the production process \( e^- p \to \nu_e H^- j \).The neutral Higgs bosons in the BSM scenario are denoted by \( \Phi_i^0 = h, H, A \).}
    \label{fig:FD}
\end{figure}

Currently, there are no direct experimental limits on \( H^\pm \) from \( e^-p \) colliders such as the proposed LHeC or FCC-eh, as these facilities remain in the design and planning stages. However, the ATLAS and CMS collaborations at the LHC have performed extensive searches for charged Higgs bosons across various production modes and decay channels. These searches constrain the charged Higgs mass to roughly the 200~GeV to 2--3~TeV range, depending on the specific final state and production mechanism. The relevant search channels and corresponding experimental bounds are summarized in~\autoref{tab:Hclimits}. These existing constraints serve as a guideline for selecting benchmark scenarios in our study, ensuring that the chosen parameter space remains both phenomenologically viable and theoretically motivated within the considered framework. To set the stage, we first illustrate the production of the charged Higgs boson in the \( e^-p \) collider environment through the most relevant Feynman diagrams, shown in~\autoref{fig:FD}. These diagrams correspond to the process \( e^- p \to \nu_e H^- q \), where \( q \) represents both light quarks \( (u, d, c, s) \) as well as the \( b \)-quark.


\begin{table}[h!]
    \centering
    \footnotesize\setlength{\tabcolsep}{4pt}
    \begin{tabular}{!{\vrule}c!{\vrule}c!{\vrule}c!{\vrule}c!{\vrule}c!{\vrule}c!{\vrule}}
        \toprule
        \textbf{Exp.} & \textbf{Channel} & $\boldsymbol{\sqrt{s}}$ & $\boldsymbol{\mathcal{L}}$ & \textbf{Mass Range} & $\boldsymbol{\sigma_{\text{upper}}\,[\mathrm{pb}]}$ \\ 
        \midrule
        ATLAS~\cite{ATLAS:2021upq} & $\sigma(pp \rightarrow H^{\pm}tb)\,\mathcal{BR}(H^{\pm} \rightarrow tb)$ & 13 TeV & 139 fb$^{-1}$ & 200 GeV--2 TeV & 3.6--0.036 \\ 
        \midrule
        CMS~\cite{CMS:2020imj} & $\sigma(pp \rightarrow H^{\pm}t[b] + pp \rightarrow H^{\pm})\,\mathcal{BR}(H^{\pm} \rightarrow tb)$ & 13 TeV & 35.9 fb$^{-1}$ & 200 GeV--3 TeV & 21.3--0.007 \\ 
        \midrule
        ATLAS~\cite{ATLAS:2018gfm} & $\sigma(pp \rightarrow H^{\pm}tb)\,\mathcal{BR}(H^{\pm} \rightarrow \tau^{\pm}\nu_{\tau})$ & 13 TeV & 36.1 fb$^{-1}$ & 90 GeV--2 TeV & 4.2--$2.5\times10^{-3}$ \\ 
        \midrule
        ATLAS~\cite{ATLAS:2016avi} & $\sigma(pp \rightarrow H^{\pm}t[b])\,\mathcal{BR}(H^{\pm} \rightarrow \tau^{\pm}\nu_{\tau})$ & 13 TeV & 3.2 fb$^{-1}$ & 200 GeV--2 TeV & 1.9--$15\times10^{-3}$ \\ 
        \midrule
        CMS~\cite{CMS:2015lsf} & $\sigma(pp \rightarrow H^{\pm}t[b])\,\mathcal{BR}(H^{\pm} \rightarrow \tau^{\pm}\nu_{\tau})$ & 8 TeV & 19.7 fb$^{-1}$ & 200 GeV--600 GeV & 2.0--0.13 \\ 
        \midrule
        ATLAS~\cite{ATLAS:2018iui} & $\sigma(pp \rightarrow H^{\pm}jj \rightarrow W^{\pm}Zjj)$ & 13 TeV & 36.1 fb$^{-1}$ & 200 GeV--900 GeV & 0.25--0.05 \\ 
        \midrule
        CMS~\cite{CMS:2021wlt} & $\sigma(pp \rightarrow H^{\pm}jj)\,\mathcal{BR}(H^{\pm} \rightarrow W^{\pm}Z)$ & 13 TeV & 137 fb$^{-1}$ & 200 GeV--3 TeV & 0.43--0.02 \\ 
        \bottomrule
    \end{tabular}
    \caption{Summary of charged Higgs boson searches at ATLAS and CMS~\cite{Coleppa:2021wjx}, indicating the channels, center-of-mass energies, integrated luminosities, mass ranges, and corresponding upper limits on the production cross section.}
    \label{tab:Hclimits}
\end{table}

A detailed analysis of the production channels, their subsequent decays, and the resulting final states used in our study to probe the charged Higgs boson is presented in \autoref{tab:pheno}.

\subsection{Charged Higgs Channels}
\label{tab:alternatives}

We now turn to the question of listing all possible signatures of the charged Higgs boson \( H^- \) focussing on its conventional decay modes. In particular, we consider both fermionic and scalar decay channels, which yield experimentally accessible final states at future \( e^-p \) colliders:  

\begin{enumerate}
    \item \textbf{Fermionic decay:} \( H^- \to \bar{t}b \), followed by \( \bar{t} \to \bar{b}W^-\). The subsequent \( W^- \) boson decay can proceed either hadronically (\( W^- \to jj \)) or leptonically (\( W^- \to \ell^- \bar{\nu}_\ell \)), resulting in distinct event topologies.
    
    \item \textbf{Scalar decay:} \( H^- \to W^- h \), where the neutral scalar further decays to \(b\bar{b}\). Again, the \( W^- \) boson may decay hadronically or leptonically, providing complementary final-state signatures.
\end{enumerate}

Based on these decay modes, we define the following hadronic signal topologies:  

\begin{itemize}
    \item \textbf{Signal~1:} \( e^- p \to \nu_e H^- j \),  
    \( H^- \to \bar{t} b \), \( \bar{t} \to \bar{b} W^- \), \( W^- \to jj \)  
    \(\Rightarrow 2b + 3j + \cancel{E}_T \)
    
    \item \textbf{Signal~2:} \( e^- p \to \nu_e H^- j \),  
    \( H^- \to W^- h \), \( W^- \to jj \), \( h \to b\bar{b} \)  
    \(\Rightarrow 2b + 3j + \cancel{E}_T \)
\end{itemize}

An important feature of Signal~1 and Signal~2 is that, despite arising from different intermediate decay chains, they both lead to the same hadronic final state: \( 2b + 3j + \cancel{E}_T \). In addition to these hadronic channels, we also consider two complementary leptonic scenarios, where the intermediate \( W^- \) boson decays leptonically. These channels produce cleaner signatures with isolated charged leptons in the final state, offering better prospects for background suppression. The corresponding leptonic signals are:  

\begin{itemize}
    \item \textbf{Signal~3:} \( e^- p \to \nu_e H^- j \),  
    \( H^- \to \bar{t} b \), \( \bar{t} \to \bar{b} W^- \), \( W^- \to \ell^- \bar{\nu}_\ell \)  
    \(\Rightarrow 2b + j + \ell^- + \cancel{E}_T \)
    
    \item \textbf{Signal~4:} \( e^- p \to \nu_e H^- j \),  
    \( H^- \to W^- h \), \( W^- \to \ell^- \bar{\nu}_\ell \), \( h \to b\bar{b} \)  
    \(\Rightarrow 2b + j + \ell^- + \cancel{E}_T \)
\end{itemize}

Thus, Signals~3 and~4 also share the same leptonic final-state topology, despite originating from distinct intermediate decays. With these four signal scenarios identified, we now proceed to analyze their discovery potential at the LHeC and FCC-eh, focusing on the feasibility of achieving a \( 5\sigma \) significance\footnote{For the exotic decay channel \( H^- \to W' Z \), no viable discovery sensitivity could be achieved at either the LHeC or FCC-eh. For completeness, the details are provided in Appendix~\ref{app}.}. The detailed collider study and the results of the corresponding phenomenological analyses for these signals are presented in~\autoref{tab:pheno}.

\section{Phenomenological Analysis}
\label{tab:pheno}

In this section, we evaluate the feasibility of achieving a \(5\sigma\) discovery of the charged Higgs boson across the various signal channels introduced in the previous sections. The signal and background event samples are generated using \texttt{MadGraph5\_aMC@NLO}~\cite{Alwall:2014hca}, followed by parton showering and hadronization with \texttt{PYTHIA8}~\cite{Sjostrand:2006za}. Detector-level effects are simulated with \texttt{Delphes3}~\cite{deFavereau:2013fsa}, employing an \(e^-p\)-collider–specific detector configuration. The BSM interactions considered in this work are implemented via a UFO model generated using \texttt{FeynRules}~\cite{Christensen:2009jx,Alloul:2013bka}. The analysis is organized according to the final-state topologies of the signal processes. For convenience, we list below the considered signal scenarios along with their respective sections:  

\begin{itemize}
    \item \textbf{Signal~1}~(\autoref{tab:hptojj}):  
    \( e^- p \to \nu_e H^- j \),  
    \( H^- \to \bar{t} b \), \( \bar{t} \to \bar{b} W^- \), \( W^- \to jj \)  
    \(\Rightarrow 2b + 3j + \cancel{E}_T \).
    
    \item \textbf{Signal~2}~(\autoref{tab:hptojj}):  
    \( e^- p \to \nu_e H^- j \),  
    \( H^- \to W^- h \), \( W^- \to jj \), \( h \to b\bar{b} \)  
    \(\Rightarrow 2b + 3j + \cancel{E}_T \).

    \item \textbf{Signal~3}~(\autoref{tab:hptolep}):  
    \( e^- p \to \nu_e H^- j \),  
    \( H^- \to \bar{t} b \), \( \bar{t} \to \bar{b} W^- \), \( W^- \to \ell^- \bar{\nu}_\ell \)  
    \(\Rightarrow 2b + j + \ell^- + \cancel{E}_T \).
        
    \item \textbf{Signal~4}~(\autoref{tab:hptolep}):  
    \( e^- p \to \nu_e H^- j \),  
    \( H^- \to W^- h \), \( W^- \to \ell^- \bar{\nu}_\ell \), \( h \to b\bar{b} \)  
    \(\Rightarrow 2b + j + \ell^- + \cancel{E}_T \).
\end{itemize}

With this setup, we aim to optimize the discovery prospects of each signal channel by designing suitable kinematic selection criteria that effectively suppress SM backgrounds while retaining a significant fraction of the signal. The SM backgrounds are initially treated in a fiducial manner, as our primary goal is to evaluate the cut efficiencies. This allows us to later rescale the background cross sections accordingly after the selection cuts. Model-specific assumptions and the benchmark parameters relevant to the BSM framework are summarized in \autoref{tab:PI}.

\subsection{Signal 1 \& 2: \( 2b + 3j + \cancel{E}_T \) Final States from Fermionic and Scalar Decays}  
\label{tab:hptojj}

Signal~1 and Signal~2 arise from two well-motivated decay modes of the charged Higgs boson,  
\[
H^- \to \bar{t} b \quad \text{and} \quad H^- \to W^- h \, ,
\]  
with subsequent decays ultimately producing the same final-state topology: \( 2b + 3j + \cancel{E}_T \). Since both signals yield identical reconstructed signatures, they also share the same dominant SM backgrounds, summarized in \autoref{tab:sig2sig3SMBG}. The signal and background samples are simulated under both the LHeC and FCC-eh collider setups for a consistent evaluation. To suppress background contamination, we first impose basic object multiplicity requirements: $N(b) \geq 2 $ and $ N(j) \geq 3$. Beyond these baseline cuts, optimized kinematic selections are applied to enhance the signal significance. The detailed cutflow charts and selection strategies for Signal~1 at both the LHeC and FCC-eh are presented in \autoref{s1:lhecFCCeh}. Likewise, the corresponding analyses for Signal~2 at both colliders are provided in \autoref{s2:LHeCFCCeh}.

\begin{table}[h]
    \centering
    \begin{tabular}{ !{\vrule}c !{\vrule} c !{\vrule} c !{\vrule} c !{\vrule} c !{\vrule}} 
        \toprule
        \textbf{BG Label} & \textbf{Background Process} & \textbf{Final State} & \textbf{\(\sigma_{\text{LHeC}}\) [pb]} & \textbf{\(\sigma_{\text{FCC-eh}}\) [pb]} \\ 
        \midrule
        BG1 & \( e^-p \rightarrow \nu_e\, \bar{t}b\, j \) & \( 3j + 2b + \cancel{E}_T \) & $1.7\times10^{-1}$ & $2.32$ \\ 
        \midrule
        BG2 & \( e^-p \rightarrow \nu_e\, W^-\, b\bar{b}\, j \) & \( 3j + 2b + \cancel{E}_T \) & $3.146 \times 10^{-1}$ & $3.762$ \\
        \midrule
        BG3 & \( e^-p \rightarrow \nu_e\, Z\, jjj,\ Z \rightarrow b\bar{b} \) & \( 3j + 2b + \cancel{E}_T \) & $2.143 \times 10^{-3}$ & $1.8\times10^{-2}$ \\
        \midrule
        BG4 & \( e^-p \rightarrow \nu_e\, h\, jjj,\ h \rightarrow b\bar{b} \) & \( 3j + 2b + \cancel{E}_T \) & $6.227\times10^{-4}$ & $5.839\times10^{-3}$ \\
        \midrule
        BG5 & \( e^-p \rightarrow \nu_e\, t\bar{t}\, j \) & \( 5j + 2b + \cancel{E}_T \) & $1.049\times10^{-5}$ & $2.783\times10^{-4}$ \\
        \bottomrule
    \end{tabular}
    \caption{Relevant SM background processes for charged Higgs signals with the \(2b + 3j + \cancel{E}_T\) topology. The last two columns list the corresponding cross sections at the LHeC and FCC-eh collider setups.}
    \label{tab:sig2sig3SMBG}
\end{table}

In the following subsections, we analyze the cutflow and extract the corresponding cut efficiencies for both the signal and background processes based on the designed selection strategy. For each signal at a given collider setup, we consider two benchmark points: \textbf{BP--1}, with \( M_{H^\pm} = 300~\mathrm{GeV} \), and \textbf{BP--2}, with \( M_{H^\pm} = 500~\mathrm{GeV} \). Using these benchmark scenarios, we then proceed to discuss the key kinematic features of the signals at the LHeC and FCC-eh.

\subsubsection{Signal~1: LHeC — 60 $\times$ 7000~GeV (\( \sqrt{s} = 1.3~\mathrm{TeV} \)) and FCC-eh — 60 $\times$ 50000~GeV (\( \sqrt{s} = 3.5~\mathrm{TeV} \))}  
\label{s1:lhecFCCeh}  

\begin{figure}[h]
    \centering
    \includegraphics[width=0.40\linewidth]{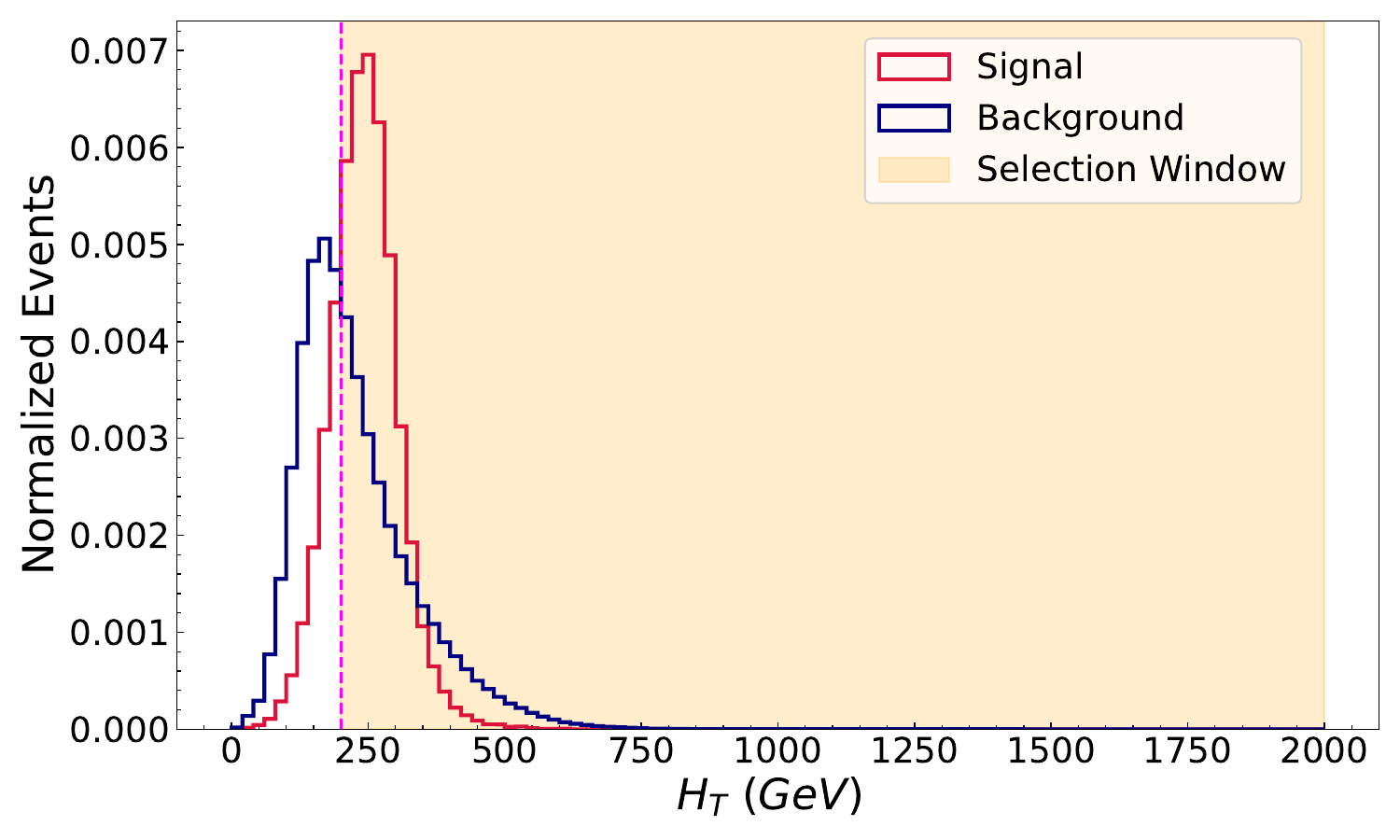}
    \caption{Distributions of kinematic observables for Signal~1 at benchmark point BP--1 (\( M_{H^\pm} = 300~\mathrm{GeV} \)). These observables are used to construct selection cuts aimed at enhancing signal significance over SM backgrounds for the final state \( 2b + 3j + \cancel{E}_T \).}
    \label{fig:sig1bp1}
\end{figure}

\begin{figure}[h]
    \centering
    \includegraphics[width=0.40\linewidth]{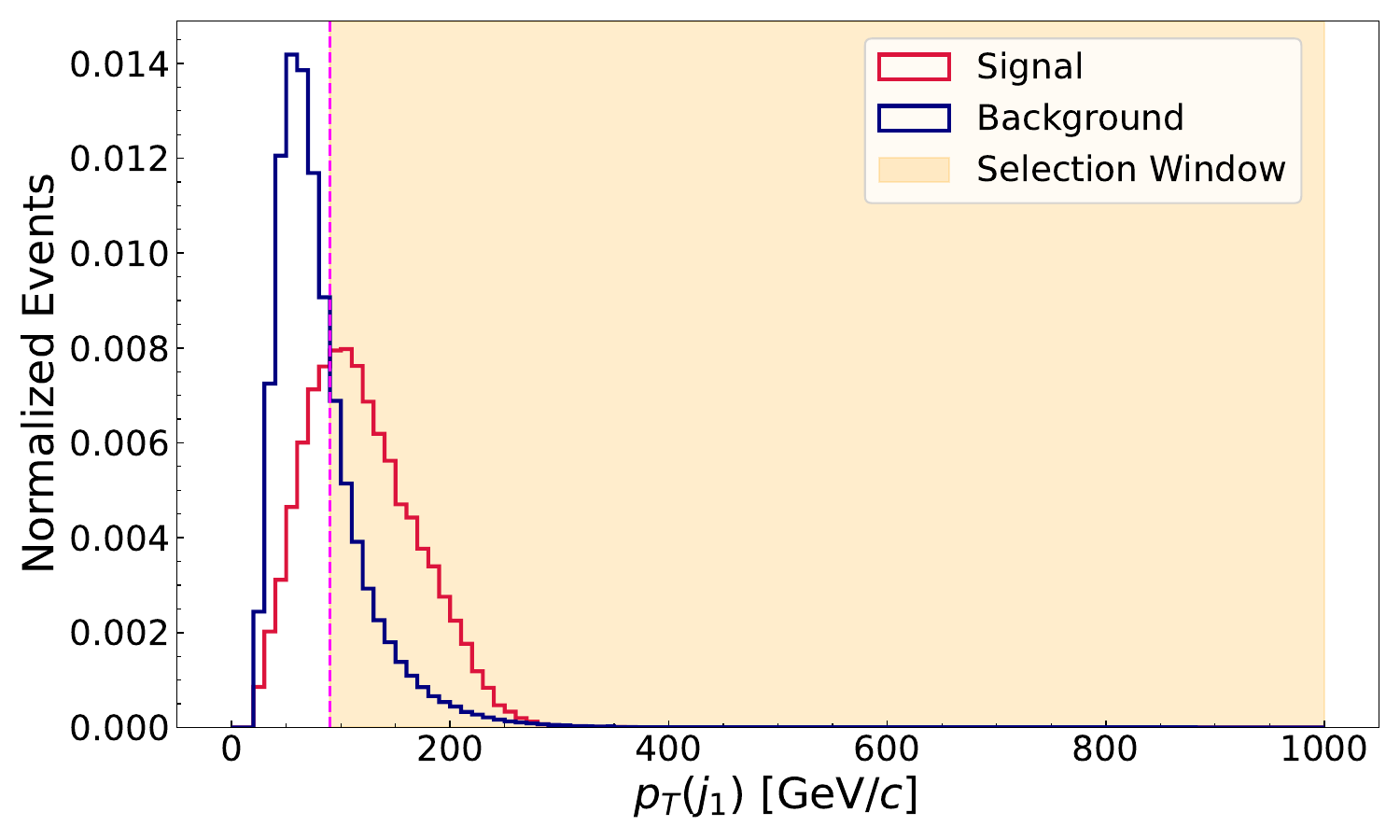}
    \includegraphics[width=0.40\linewidth]{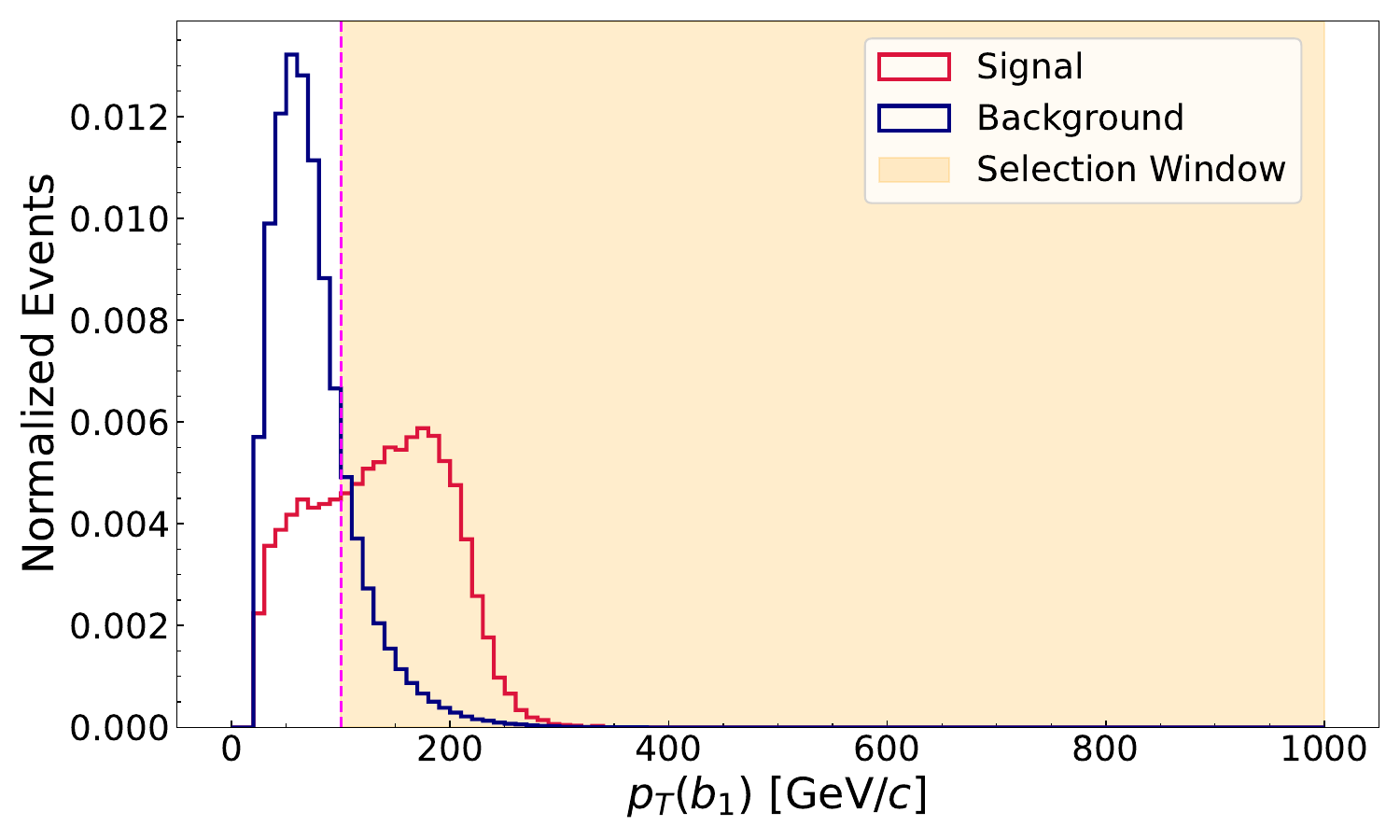}
    \includegraphics[width=0.40\linewidth]{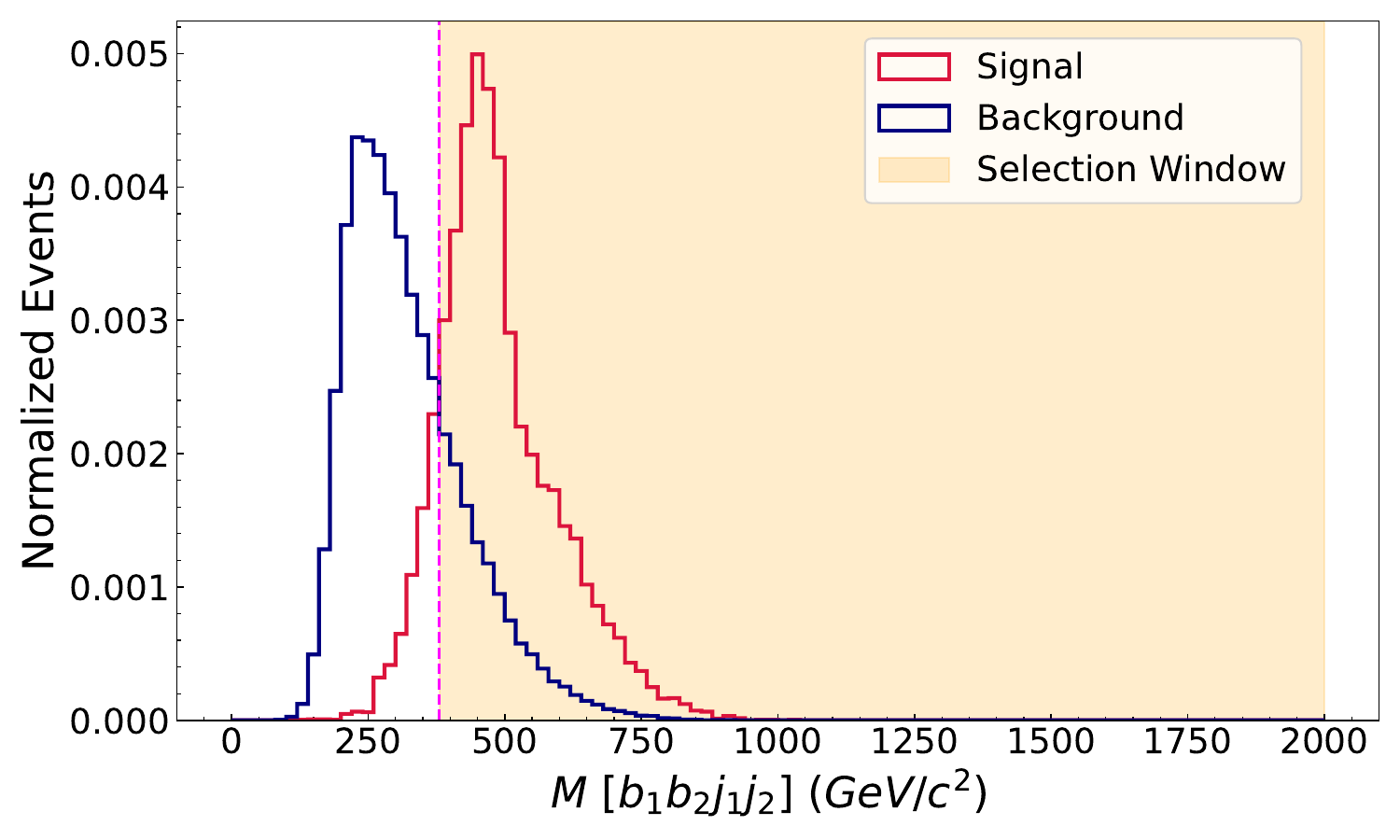}
    \includegraphics[width=0.40\linewidth]{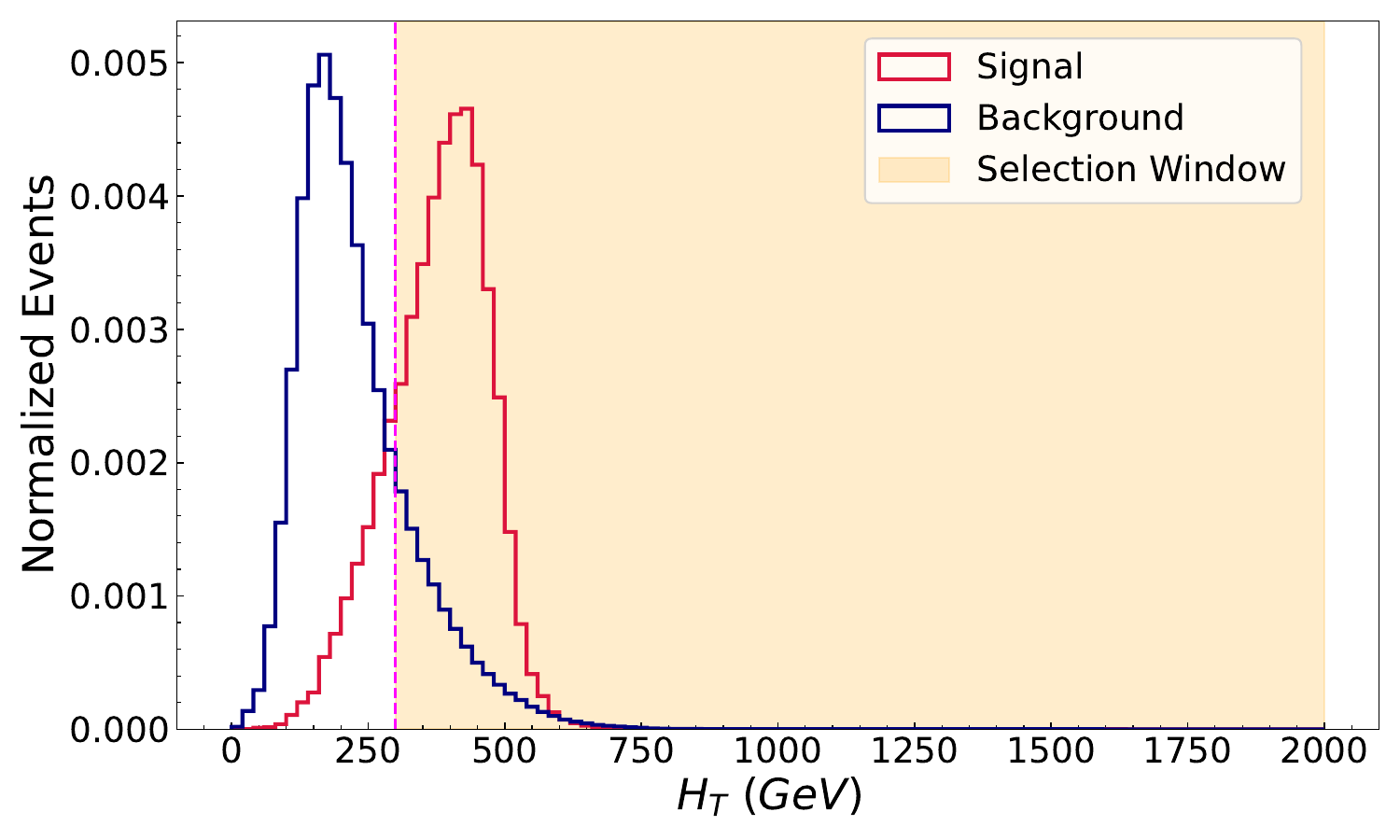}
    \caption{Distributions of kinematic observables for Signal~1 at benchmark point BP--2 (\( M_{H^\pm} = 500~\mathrm{GeV} \)). These observables are used to construct selection cuts aimed at enhancing signal significance over SM backgrounds for the final state \( 2b + 3j + \cancel{E}_T \).}
    \label{fig:sig1bp2}
\end{figure}

Signal~1, as previously discussed, results in the final-state topology \( 2b + 3j + \cancel{E}_T \), which naturally motivates the application of basic object multiplicity requirements, namely \( N(b) \geq 2 \) and \( N(j) \geq 3 \). A detailed examination of the event kinematics reveals additional observables that are effective for improving signal sensitivity while suppressing the backgrounds. In particular, the transverse momentum of the leading jet (\( p_T(j_1) \)) and the leading \( b \)-jet (\( p_T(b_1) \)), the invariant mass of the two leading \( b \)-jets and two leading jets (\( M_{b_1 b_2 j_1 j_2} \)), and the scalar sum of transverse hadronic energy (\( H_T \)) are identified as key discriminating variables. The corresponding kinematic distributions for these observables, which guide the design of the optimized cutflow, are shown in \autoref{fig:sig1bp1} for \textbf{BP--1} and in \autoref{fig:sig1bp2} for \textbf{BP--2}. The same strategy is applied consistently for both the LHeC and FCC-eh collider setups\footnote{The signal distributions at $\sqrt{s}=1.5~\mathrm{TeV}$ (LHeC) and 
$\sqrt{s}=3.5~\mathrm{TeV}$ (FCC-eh) were explicitly compared and found to have 
nearly identical shapes, as the kinematics are driven mainly by the fixed 
$m_{H^-}$ resonance rather than the collider energy. Hence, the same cutflow is 
used for both setups for consistency, with the main difference being the total yield.} to ensure a unified approach in evaluating the discovery potential of Signal~1. For the LHeC setup, the resulting cutflow chart is presented in \autoref{tab:sig1-2_LHeC_combined}, while the corresponding results for the FCC-eh setup are summarized in \autoref{tab:sig1-2_FCC-eh_combined}.

\begin{table}[t]
\centering
\begin{minipage}{0.50\linewidth} 
\centering
\begin{tabular}{!{\vrule}c!{\vrule}c!{\vrule}c!{\vrule}c!{\vrule}c!{\vrule}c!{\vrule}c!{\vrule}}
\toprule
\multicolumn{7}{!{\vrule}c!{\vrule}}{\textbf{(a) BP--1: \(M_{H^\pm}=300~\mathrm{GeV}\)}} \\
\midrule
\textbf{Cuts} & \textbf{S} & \textbf{BG1} & \textbf{BG2} & \textbf{BG3} & \textbf{BG4} & \textbf{BG5} \\
\midrule
-- & 100000 & 300000 & 400000 & 100000 & 100000 & 100000 \\
\( N_b \ge 2 \) & 28121 & 51420 & 52775 & 20728 & 27469 & 22749 \\
\( N_j \ge 3 \) & 12081 & 18703 & 15711 & 7319.8 & 9527.2 & 19805 \\
$H_T\ge200$ & 11714 & 17196 & 14245 & 6811.8 & 8974.8 & 19777 \\
\bottomrule
\end{tabular}
\end{minipage}%
\hfill
\begin{minipage}{0.50\linewidth} 
\centering
\begin{tabular}{!{\vrule}c!{\vrule}c!{\vrule}c!{\vrule}c!{\vrule}c!{\vrule}c!{\vrule}c!{\vrule}}
\toprule
\multicolumn{7}{!{\vrule}c!{\vrule}}{\textbf{(b) BP--2: \(M_{H^\pm}=500~\mathrm{GeV}\)}} \\
\midrule
\textbf{Cuts} & \textbf{S} & \textbf{BG1} & \textbf{BG2} & \textbf{BG3} & \textbf{BG4} & \textbf{BG5} \\
\midrule
-- & 100000 & 300000 & 400000 & 100000 & 100000 & 100000 \\
\( N_b \ge 2 \) & 28069 & 51420 & 52775 & 20728 & 27469 & 22749 \\
\( N_j \ge 3 \) & 14820 & 18703 & 15711 & 7319.8 & 9527.2 & 19805 \\
$p_T(j_1)\ge90$ & 7892.9 & 4017.9 & 3069.8 & 3638.0 & 3267.3 & 11972 \\
$p_T(b_1)\ge100$ & 6201.3 & 1437.7 & 1093.9 & 1116.1 & 1292.3 & 5256.1 \\
$M_{b_1b_2j_1j_2}\ge380$ & 5828.5 & 1094.4 & 845.9 & 934.1 & 1064.7 & 4380.1 \\
$H_T\ge300$ & 5822.4 & 1094.4 & 841.9 & 934.1 & 1059.8 & 4380.1 \\
\bottomrule
\end{tabular}
\end{minipage}
\caption{Cutflow results for Signal~1 at the LHeC for two benchmark points: BP--1 (\(M_{H^\pm}=300~\mathrm{GeV}\)) and BP--2 (\(M_{H^\pm}=500~\mathrm{GeV}\)).}
\label{tab:sig1-2_LHeC_combined}
\end{table}

\begin{table}[t]
\centering
\begin{minipage}{0.50\linewidth} 
\centering
\begin{tabular}{!{\vrule}c!{\vrule}c!{\vrule}c!{\vrule}c!{\vrule}c!{\vrule}c!{\vrule}c!{\vrule}}
\toprule
\multicolumn{7}{!{\vrule}c!{\vrule}}{\textbf{(a) BP--1: \(M_{H^\pm}=300~\mathrm{GeV}\)}} \\
\midrule
\textbf{Cuts} & \textbf{S} & \textbf{BG1} & \textbf{BG2} & \textbf{BG3} & \textbf{BG4} & \textbf{BG5} \\
\midrule
-- & 100000 & 300000 & 400000 & 100000 & 100000 & 100000 \\
\( N_b \ge 2 \) & 21780 & 39676.1 & 43061.6 & 16703 & 23190 & 8146.6 \\
\( N_j \ge 3 \) & 10475.3 & 17976.5 & 17277.3 & 8475.2 & 11628 & 7371.6 \\
$H_T\ge200$ & 10236.3 & 17252.6 & 16411.1 & 8239.7 & 11299 & 7366.6 \\
\bottomrule
\end{tabular}
\end{minipage}%
\hfill
\begin{minipage}{0.50\linewidth} 
\centering
\begin{tabular}{!{\vrule}c!{\vrule}c!{\vrule}c!{\vrule}c!{\vrule}c!{\vrule}c!{\vrule}c!{\vrule}}
\toprule
\multicolumn{7}{!{\vrule}c!{\vrule}}{\textbf{(b) BP--2: \(M_{H^\pm}=500~\mathrm{GeV}\)}} \\
\midrule
\textbf{Cuts} & \textbf{S} & \textbf{BG1} & \textbf{BG2} & \textbf{BG3} & \textbf{BG4} & \textbf{BG5} \\
\midrule
-- & 100000 & 300000 & 400000 & 100000 & 100000 & 100000 \\
\( N_b \ge 2 \) & 18482 & 39676.1 & 43061.6 & 16703 & 23190 & 8146.6 \\
\( N_j \ge 3 \) & 11304 & 17976.5 & 17277.3 & 8475.2 & 11628 & 7371.6 \\
$p_T(j_1)\ge90$ & 7142.8 & 6437.3 & 5905.4 & 5730.4 & 6188.5 & 5751.6 \\
$p_T(b_1)\ge100$ & 5552.4 & 3222.5 & 2968.0 & 2397.2 & 3220.6 & 3222.0 \\
$M_{b_1b_2j_1j_2}\ge380$ & 5343.8 & 2911.7 & 2674.2 & 2269.8 & 3105.4 & 3018.0 \\
$H_T\ge300$ & 5341.8 & 2910.8 & 2673.0 & 2267.1 & 3097.7 & 3018.0 \\
\bottomrule
\end{tabular}
\end{minipage}
\caption{Cutflow results for Signal~1 at the FCC-eh for two benchmark points: BP--1 (\(M_{H^\pm}=300~\mathrm{GeV}\)) and BP--2 (\(M_{H^\pm}=500~\mathrm{GeV}\)).}
\label{tab:sig1-2_FCC-eh_combined}
\end{table}

\subsubsection{Signal~2: LHeC — 60 $\times$ 7000~GeV (\( \sqrt{s} = 1.3~\mathrm{TeV} \)) and FCC-eh — 60 $\times$ 50000~GeV (\( \sqrt{s} = 3.5~\mathrm{TeV} \))}  
\label{s2:LHeCFCCeh}  

Similar to Signal~1, we analyze the kinematic features of Signal~2 to design optimized selection cuts and construct the corresponding cutflow charts. The same strategy is applied for both the LHeC and FCC-eh collider setups to ensure consistency in evaluating the signal sensitivity and background suppression. The resulting cutflow chart for Signal~2 in the LHeC setup is presented in \autoref{tab:s2LHec}, while the corresponding results for the FCC-eh setup are summarized in \autoref{tab:s2FCCeh}.

\begin{table}[t]
\centering
\begin{minipage}{0.50\linewidth} 
\centering
\begin{tabular}{!{\vrule}c!{\vrule}c!{\vrule}c!{\vrule}c!{\vrule}c!{\vrule}c!{\vrule}c!{\vrule}}
\toprule
\multicolumn{7}{!{\vrule}c!{\vrule}}{\textbf{(a) BP--1: \(M_{H^\pm}=300~\mathrm{GeV}\)}} \\
\midrule
\textbf{Cuts} & \textbf{S} & \textbf{BG1} & \textbf{BG2} & \textbf{BG3} & \textbf{BG4} & \textbf{BG5} \\
\midrule
        -- & 100000 & 300000 & 400000 & 100000 & 100000 & 100000 \\
        
        \( N_b \geq 2 \) & 26077 & 51420 & 52775 & 20728 & 27469 & 22749 \\
        
        \( N_j \geq 3 \) & 11645 & 18703 & 15711 & 7319.8 & 9527.2 & 19805 \\
        
        $H_T\geq200$ & 11273 & 17196 & 14245 & 6811.8 & 8974.8 & 19777 \\
        \bottomrule
\end{tabular}
\end{minipage}%
\hfill
\begin{minipage}{0.50\linewidth} 
\centering
\begin{tabular}{!{\vrule}c!{\vrule}c!{\vrule}c!{\vrule}c!{\vrule}c!{\vrule}c!{\vrule}c!{\vrule}}
\toprule
\multicolumn{7}{!{\vrule}c!{\vrule}}{\textbf{(b) BP--2: \(M_{H^\pm}=500~\mathrm{GeV}\)}} \\
\midrule
\textbf{Cuts} & \textbf{S} & \textbf{BG1} & \textbf{BG2} & \textbf{BG3} & \textbf{BG4} & \textbf{BG5} \\
\midrule
        -- & 100000 & 300000 & 400000 & 100000 & 100000 & 100000 \\
        
        \( N_b \geq 2 \) & 27011 & 51420 & 52775 & 20728 & 27469 & 22749 \\
        
        \( N_j \geq 3 \) & 15043 & 18703 & 15711 & 7319.8 & 9527.2 & 19805 \\

        $P_T(j_1)\geq100$ & 11389 & 2791.3 & 2028.0 & 3104.0 & 2554.8 & 10174.1 \\
        
        $P_T(b_1)\geq100$ & 9182.8 & 1096.8 & 783.9 & 1010.0 & 1069.8 & 4512.1 \\

        $\Delta R(b_1,b_2)\leq1.6$ & 7602.7 & 326.5 & 228.0 & 814.1 & 682.4 & 1600.0 \\
        
        $M_{b_1b_2j_1j_2}\geq400$ & 7381.0 & 201.7 & 142.0 & 670.1 & 599.9 & 1284.0 \\
        
        $H_T\geq300$ & 7381.0 & 201.7 & 142.0 & 670.1 & 599.9 & 1284.0 \\
        \bottomrule
\end{tabular}
\end{minipage}
\caption{Cutflow results for Signal~2 at the LHeC for two benchmark points: BP--1 (\(M_{H^\pm}=300~\mathrm{GeV}\)) and BP--2 (\(M_{H^\pm}=500~\mathrm{GeV}\)).}
\label{tab:s2LHec}
\end{table}

\begin{table}[t]
\centering
\begin{minipage}{0.50\linewidth} 
\centering
\begin{tabular}{!{\vrule}c!{\vrule}c!{\vrule}c!{\vrule}c!{\vrule}c!{\vrule}c!{\vrule}c!{\vrule}}
\toprule
\multicolumn{7}{!{\vrule}c!{\vrule}}{\textbf{(a) BP--1: \(M_{H^\pm}=300~\mathrm{GeV}\)}} \\
\midrule
\textbf{Cuts} & \textbf{S} & \textbf{BG1} & \textbf{BG2} & \textbf{BG3} & \textbf{BG4} & \textbf{BG5} \\
\midrule
        -- & 100000 & 300000 & 400000 & 100000 & 100000 & 100000 \\
        
        \( N_b \geq 2 \) & 20227 & 39676 & 43061 & 16703 & 23190 & 8146.6 \\
        
        \( N_j \geq 3 \) & 9894.6 & 17976 & 17277 & 8475.2 & 11628 & 7371.6 \\
    
        $H_T\geq200$ & 9633.9 & 17252 & 16411 & 8239.7 & 11299 & 7366.6 \\
        \bottomrule
\end{tabular}
\end{minipage}%
\hfill
\begin{minipage}{0.50\linewidth} 
\centering
\begin{tabular}{!{\vrule}c!{\vrule}c!{\vrule}c!{\vrule}c!{\vrule}c!{\vrule}c!{\vrule}c!{\vrule}}
\toprule
\multicolumn{7}{!{\vrule}c!{\vrule}}{\textbf{(b) BP--2: \(M_{H^\pm}=500~\mathrm{GeV}\)}} \\
\midrule
\textbf{Cuts} & \textbf{S} & \textbf{BG1} & \textbf{BG2} & \textbf{BG3} & \textbf{BG4} & \textbf{BG5} \\
\midrule
-- & 100000 & 300000 & 400000 & 100000 & 100000 & 100000 \\
        
        \( N_b \geq 2 \) & 18682 & 39676 & 43061 & 16703 & 23190 & 8146.6 \\

        \( N_j \geq 3 \) & 11635 & 17976 & 17277 & 8475.2 & 11628 & 7371.6 \\
 
        $P_T(j_1)\geq100$ & 8880.2 & 5110.5 & 4640.3 & 5182.9 & 5371.1 & 5241.6 \\

        $P_T(b_1)\geq100$ & 6493.6 & 2671.8 & 2459.6 & 2233.9 & 2910.5 & 2977.0 \\

        $\Delta R(b_1,b_2)\leq1.6$ & 5200.0 & 800.1 & 760.3 & 1709.6 & 2043.4 & 1227.9 \\
 
        $M_{b_1b_2j_1j_2}\geq400$ & 5083.4 & 687.9 & 656.0 & 1607.8 & 1986.9 & 1124.0 \\

        $H_T\geq300$ & 5083.4 & 687.9 & 656.0 & 1607.8 & 1986.9 & 1124.0 \\
        \bottomrule
\end{tabular}
\end{minipage}
\caption{Cutflow results for Signal~2 at the FCC-eh for two benchmark points: BP--1 (\(M_{H^\pm}=300~\mathrm{GeV}\)) and BP--2 (\(M_{H^\pm}=500~\mathrm{GeV}\)).}
\label{tab:s2FCCeh}
\end{table}

\subsection{Signal~3 \& 4: \( 2b + j + \ell^- + \cancel{E}_T \) Final States from Fermionic and Scalar Decays}  
\label{tab:hptolep}  

Signal~3 and Signal~4 are analogous to Signal~1 and Signal~2, with the key difference being that the intermediate \( W^- \) boson in the decay chain undergoes a leptonic decay. This results in a distinct final-state topology featuring an isolated charged lepton, which can significantly improve the ability to discriminate signal events from the SM backgrounds. As in the previous cases, we consider two benchmark points: \textbf{BP--1} with \( M_{H^\pm} = 300~\mathrm{GeV} \) and \textbf{BP--2} with \( M_{H^\pm} = 500~\mathrm{GeV} \). The relevant SM background processes for these leptonic final states, along with their corresponding cross sections for both the LHeC and FCC-eh setups, are summarized in \autoref{tab:case2SMBG}. With these SM backgrounds defined, we now examine the kinematic features of Signal~3 and Signal~4 in both collider environments.

\begin{table}[h]
    \centering
    \begin{tabular}{ !{\vrule}c !{\vrule} c !{\vrule} c !{\vrule} c !{\vrule} c !{\vrule}} 
        \toprule
        \textbf{Background Label} & \textbf{Background Process} & \textbf{Final State} & \textbf{$\sigma_{\text{LHeC}}$ [pb]} & \textbf{$\sigma_{\text{FCC-eh}}$ [pb]} \\ 
        \midrule
        BG1 & $e^-p \rightarrow \nu_e\, t\bar{b}\, j$ & $2b + \ell^- + j + \cancel{E}_T$ & $5.663\times10^{-2}$ & $7.743\times10^{-1}$ \\ 
        \midrule
        BG2 & $e^-p \rightarrow \nu_e\, W^-\, b\bar{b}\, j$ & $2b + \ell^- + j + \cancel{E}_T$ & $1.049\times10^{-1}$ & $1.253$ \\
        \midrule
        BG3 & $e^-p \rightarrow \nu_e\, Z\, jjj,\, Z \rightarrow b\bar{b}$ & $2b + \ell^- + j + \cancel{E}_T$ & $2.144\times10^{-3}$ & $1.888\times10{-2}$ \\
        \midrule
        BG4 & $e^-p \rightarrow \nu_e\, h\, jjj,\, h \rightarrow b\bar{b}$ & $2b + \ell^- + j + \cancel{E}_T$ & $6.227\times10^{-4}$ & $5.839\times10^{-3}$ \\
        \midrule
        BG5 & $e^-p \rightarrow \nu_e\, t\bar{t}\, j$ & $5j + 2b + \cancel{E}_T$ & $1.049\times10^{-5}$ & $2.783\times10^{-4}$ \\
        \bottomrule
    \end{tabular}
    \caption{Relevant SM background processes for charged Higgs signal processes with final state \( \ell^- + j + 2b + \cancel{E}_T \), along with their cross sections at the LHeC and FCC-eh.}
    \label{tab:case2SMBG}
\end{table}

\subsubsection{Signal 3: LHeC — 60 $\times$ 7000 GeV ($\sqrt{s} = 1.3~\mathrm{TeV}$) $\&$ FCC-eh — 60 $\times$ 50000 GeV ($\sqrt{s} = 3.5~\mathrm{TeV}$)}

\begin{figure}[h]
    \centering
    \includegraphics[width=0.40\linewidth]{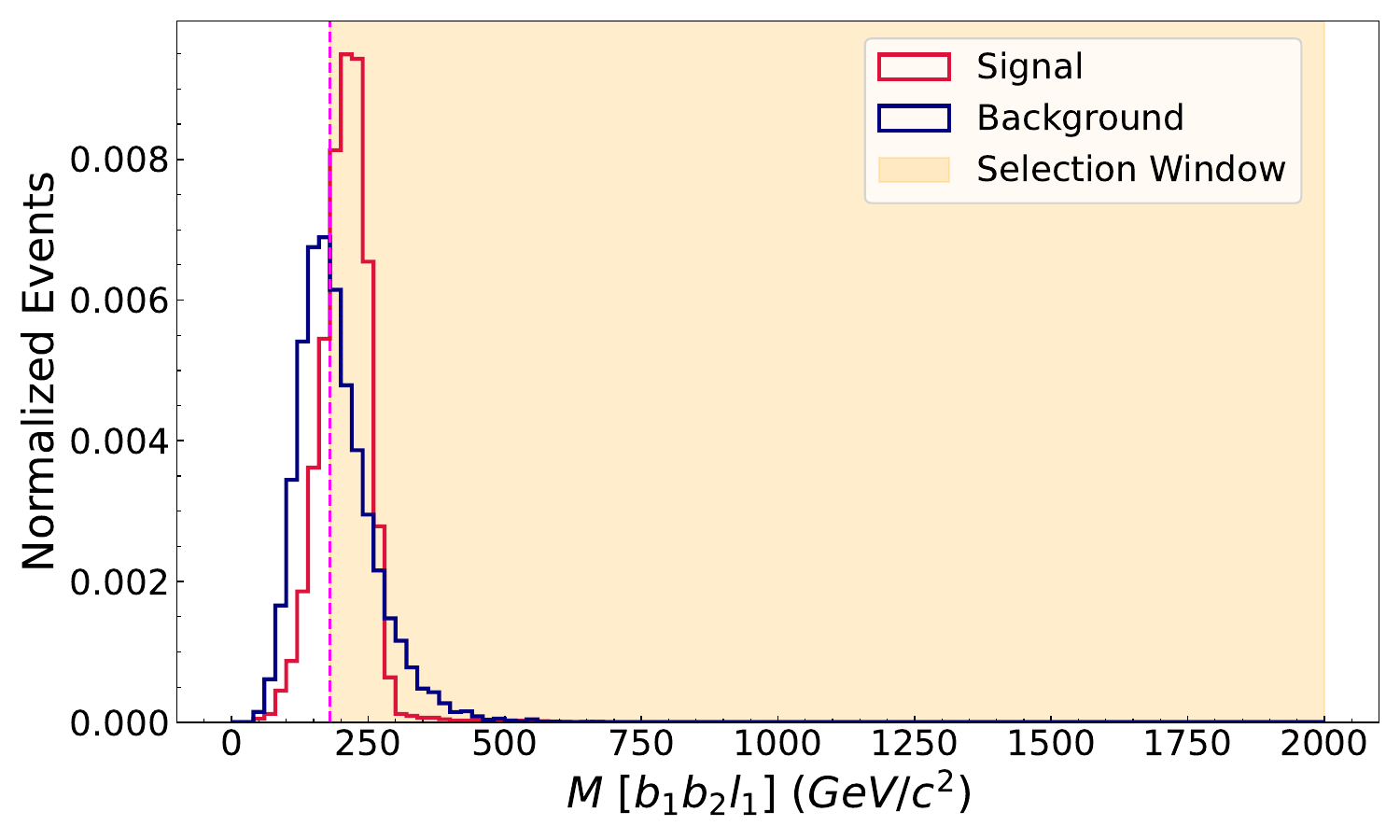}
    \includegraphics[width=0.40\linewidth]{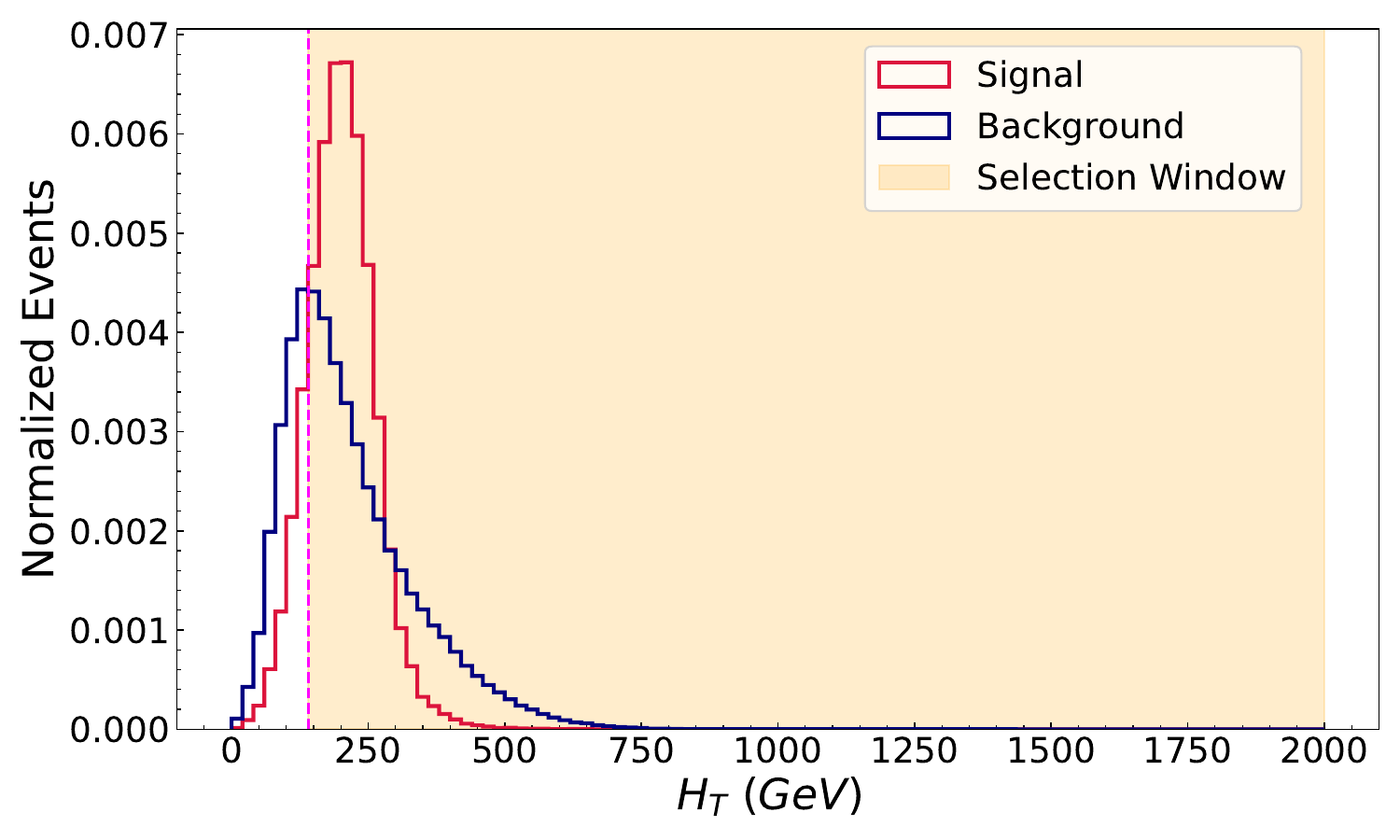}
    \caption{Distributions of kinematic observables for Signal~3 at benchmark point BP--1 (\( M_{H^\pm} = 300~\mathrm{GeV} \)). These observables are used to construct selection cuts aimed at enhancing signal significance over SM backgrounds for the final state \( 2b + \ell^- + j + \cancel{E}_T \).}
    \label{fig:sig3bp1}
\end{figure}

For the final state \( 2b + \ell^- + j + \cancel{E}_T \), we first impose the necessary object multiplicity cuts required for proper reconstruction, namely \( N(b) \geq 2 \), \( N(\ell) \geq 1 \), and \( N(j) \geq 1 \). After applying these basic selection requirements, we examined various kinematic distributions to design additional selection cuts aimed at enhancing signal sensitivity while suppressing SM backgrounds. The relevant kinematic distributions for Signal~3 in the BP--1 scenario are shown in \autoref{fig:sig3bp1}. Likewise, we identify and implemented the cutflowchart for BP-2 and the combined cuflowcharts for both benchmark points is presented in \autoref{tab:sig3LHec:combained}. Similar to that of the LHeC setup, we have carried out implementation of selection cuts in the FCC-eh setup as well and the cutflow chart is displayed in \autoref{tab:sig3FCCeh:combained}.

\begin{table}[t]
\centering
\begin{minipage}{0.48\linewidth} 
\centering
\begin{tabular}{!{\vrule}c!{\vrule}c!{\vrule}c!{\vrule}c!{\vrule}c!{\vrule}c!{\vrule}c!{\vrule}}
\toprule
\multicolumn{7}{!{\vrule}c!{\vrule}}{\textbf{(a) BP--1: \(M_{H^\pm}=300~\mathrm{GeV}\)}} \\
\midrule
\textbf{Cuts} & \textbf{S} & \textbf{BG1} & \textbf{BG2} & \textbf{BG3} & \textbf{BG4} & \textbf{BG5} \\
\midrule
        -- & 100000 & 200000 & 100000 & 100000 & 100000 & 100000 \\

        \( N_b \geq 2 \) & 24509 & 26163 & 9540.5 & 20490 & 27134 & 22844 \\

        \( N_j \geq 1 \) & 19259 & 18974 & 6244.1 & 20015 & 26617 & 22816 \\
   
        \( N_\ell \geq 1 \) & 10807.5 & 10603 & 3360.3 & 2.50 & 7.50 & 10.0 \\

        $M_{b_1b_2\ell_1}\geq180$ & 7970.2 & 5144.2 & 1580.1 & 0 & 5.0 & 2.49 \\

        $H_T\geq140$ & 7964.1 & 5120.2 & 1573.4 & 0 & 5.0 & 2.49 \\
        \bottomrule
\end{tabular}
\end{minipage}%
\hfill
\begin{minipage}{0.48\linewidth} 
\centering
\begin{tabular}{!{\vrule}c!{\vrule}c!{\vrule}c!{\vrule}c!{\vrule}c!{\vrule}c!{\vrule}c!{\vrule}}
\toprule
\multicolumn{7}{!{\vrule}c!{\vrule}}{\textbf{(b) BP--2: \(M_{H^\pm}=500~\mathrm{GeV}\)}} \\
\midrule
\textbf{Cuts} & \textbf{S} & \textbf{BG1} & \textbf{BG2} & \textbf{BG3} & \textbf{BG4} & \textbf{BG5} \\
\midrule
        -- & 100000 & 200000 & 100000 & 100000 & 100000 & 100000 \\

        \( N_b \geq 2 \) & 24727 & 26163 & 9540.5 & 20490 & 27134 & 22844 \\
   
        \( N_j \geq 1 \) & 20838 & 18974 & 6244.1 & 20015 & 26617 & 22816 \\

        \( N_\ell \geq 1 \) & 9654.1 & 10603 & 3360.3 & 2.5 & 7.5 & 10.0 \\

        $P_T(b_1)\geq110$ & 8117.9 & 2019.1 & 570.0 & 0 & 5.0 & 2.51 \\

        $\Delta R(b_1,\ell_1)\leq2.5$ & 5936.1 & 860.7 & 270.0 & 0.0 & 5.0 & 2.51 \\

        $M_{b_1b_2\ell_1}\geq260$ & 5559.0 & 572.7 & 130.0 & 0.0 & 5.0 & 0.0 \\
  
        $H_T\geq240$ & 5555.0 & 569.5 & 130.0 & 0.0 & 5.0 & 0.0 \\
        \bottomrule
\end{tabular}
\end{minipage}
\caption{Cutflow results for Signal~3 at the LHeC for two benchmark points: BP--1 (\(M_{H^\pm}=300~\mathrm{GeV}\)) and BP--2 (\(M_{H^\pm}=500~\mathrm{GeV}\)).}
\label{tab:sig3LHec:combained}
\end{table}
\begin{table}[t]
\centering
\begin{minipage}{0.48\linewidth} 
\centering
\begin{tabular}{!{\vrule}c!{\vrule}c!{\vrule}c!{\vrule}c!{\vrule}c!{\vrule}c!{\vrule}c!{\vrule}}
\toprule
\multicolumn{7}{!{\vrule}c!{\vrule}}{\textbf{(a) BP--1: \(M_{H^\pm}=300~\mathrm{GeV}\)}} \\
\midrule
\textbf{Cuts} & \textbf{S} & \textbf{BG1} & \textbf{BG2} & \textbf{BG3} & \textbf{BG4} & \textbf{BG5} \\
\midrule
        -- & 100000 & 200000 & 100000 & 100000 & 100000 & 100000 \\

        \( N_b \geq 2 \) & 18860 & 20449 & 7949.5 & 16663 & 23180 & 8248.6 \\

        \( N_j \geq 1 \) & 15669 & 16061 & 5908.6 & 16480 & 22956 & 8246.6 \\

        \( N_\ell \geq 1 \) & 8254.1 & 7918.3 & 2906.8 & 3.01 & 6.65 & 3.0 \\
      
        $M_{b_1b_2\ell_1}\geq180$ & 5958.9 & 4268.0 & 1546.6 & 0.603 & 4.4 & 2.0 \\
    
        $H_T\geq140$ & 5954.5 & 4258.9 & 1543.3 & 0.603 & 4.4 & 2.0 \\
        \bottomrule
\end{tabular}
\end{minipage}%
\hfill
\begin{minipage}{0.48\linewidth} 
\centering
\begin{tabular}{!{\vrule}c!{\vrule}c!{\vrule}c!{\vrule}c!{\vrule}c!{\vrule}c!{\vrule}c!{\vrule}}
\toprule
\multicolumn{7}{!{\vrule}c!{\vrule}}{\textbf{(b) BP--2: \(M_{H^\pm}=500~\mathrm{GeV}\)}} \\
\midrule
\textbf{Cuts} & \textbf{S} & \textbf{BG1} & \textbf{BG2} & \textbf{BG3} & \textbf{BG4} & \textbf{BG5} \\
\midrule
        -- & 100000 & 200000 & 100000 & 100000 & 100000 & 100000 \\

        \( N_b \geq 2 \) & 15410 & 20049 & 7949.5 & 16663 & 23180 & 8248.6 \\

        \( N_j \geq 1 \) & 13835 & 16061 & 5908.6 & 16480 & 22956 & 8246.6 \\
 
        \( N_\ell \geq 1 \) & 5608.0 & 7918.3 & 2906.8 & 3.01 & 6.65 & 3.0 \\

        $P_T(b_1)\geq110$ & 4646.7 & 2124.2 & 782.7 & 0.601 & 3.32 & 0.997 \\
  
        $\Delta R(b_1,\ell_1)\leq2.5$ & 3127.2 & 885.8 & 330.1 & 0.0 & 1.11 & 0 \\

        $M_{b_1b_2\ell_1}\geq260$ & 2932.8 & 633.3 & 230.1 & 0.0 & 0.0 & 0.0 \\
 
        $H_T\geq240$ & 2932.8 & 628.9 & 227.6 & 0.0 & 0.0 & 0.0 \\
        \bottomrule
\end{tabular}
\end{minipage}
\caption{Cutflow results for Signal~3 at the FCC-eh for two benchmark points: BP--1 (\(M_{H^\pm}=300~\mathrm{GeV}\)) and BP--2 (\(M_{H^\pm}=500~\mathrm{GeV}\)).}
\label{tab:sig3FCCeh:combained}
\end{table}

\subsubsection{Signal 4: LHeC — 60 $\times$ 7000 GeV ($\sqrt{s} = 1.3$ TeV) $\&$ FCC-eh — 60 $\times$ 50000 GeV ($\sqrt{s} = 3.5$ TeV)}

We now move on to Signal~4, where the charged Higgs boson decays as $H^- \rightarrow W^- h$, ultimately leading to the final state \( 2b + \ell^- + j + \cancel{E}_T \). The kinematic properties of this signal were carefully examined, and optimized selection cuts were derived accordingly. The resulting cutflow charts for both benchmark points are presented for the LHeC setup in \autoref{tab:sig4LHec:combained} and for the FCC-eh setup in \autoref{tab:sig4FCCeh:combained}.

\begin{table}[t]
\centering
\begin{minipage}{0.48\linewidth} 
\centering
\begin{tabular}{!{\vrule}c!{\vrule}c!{\vrule}c!{\vrule}c!{\vrule}c!{\vrule}c!{\vrule}c!{\vrule}}
\toprule
\multicolumn{7}{!{\vrule}c!{\vrule}}{\textbf{(a) BP--1: \(M_{H^\pm}=300~\mathrm{GeV}\)}} \\
\midrule
\textbf{Cuts} & \textbf{S} & \textbf{BG1} & \textbf{BG2} & \textbf{BG3} & \textbf{BG4} & \textbf{BG5} \\
\midrule
        -- & 100000 & 200000 & 100000 & 100000 & 100000 & 100000 \\
        
        \( N_b \geq 2 \) & 21165 & 26163 & 9540.5 & 20490 & 27134 & 22844 \\
        
        \( N_j \geq 1 \) & 16565 & 18974 & 6244.1 & 20015 & 26617 & 22816 \\
        
        \( N_\ell \geq 1 \) & 9255.3 & 10603 & 3360.3 & 2.50 & 7.50 & 10.0 \\
        
        $M_{b_1b_2\ell_1}\geq160$ & 7525.7 & 6622.5 & 2033.4 & 0 & 5.0 & 2.49 \\
        
        $\cancel{H}_T\geq60$ & 6196.2 & 4326.4 & 1266.8 & 0 & 2.50 & 2.49 \\
        \bottomrule
\end{tabular}
\end{minipage}%
\hfill
\begin{minipage}{0.48\linewidth} 
\centering
\begin{tabular}{!{\vrule}c!{\vrule}c!{\vrule}c!{\vrule}c!{\vrule}c!{\vrule}c!{\vrule}c!{\vrule}}
\toprule
\multicolumn{7}{!{\vrule}c!{\vrule}}{\textbf{(b) BP--2: \(M_{H^\pm}=500~\mathrm{GeV}\)}} \\
\midrule
\textbf{Cuts} & \textbf{S} & \textbf{BG1} & \textbf{BG2} & \textbf{BG3} & \textbf{BG4} & \textbf{BG5} \\
\midrule
        -- & 100000 & 200000 & 100000 & 100000 & 100000 & 100000 \\
        
        \( N_b \geq 2 \) & 22112 & 26163 & 9540.5 & 20490 & 27134 & 22844 \\
        
        \( N_j \geq 1 \) & 17746 & 18974 & 6244.1 & 20015 & 26617 & 22816 \\
        
        \( N_\ell \geq 1 \) & 8251.3 & 10603 & 3360.3 & 2.5 & 7.5 & 10.0 \\
        
        $P_T(b_1)\geq100$ & 6953.1 & 2681.6 & 773.3 & 0 & 5.0 & 2.51 \\
        
        $P_T(\ell_1)\geq70$ & 5421.0 & 591.9 & 126.6 & 0 & 0 & 0 \\
        
        $\Delta R(b_1,b_2)\leq1.8$ & 5253.7 & 177.6 & 36.67 & 0 & 0 & 0 \\
        
        $M_{b_1\ell_1}\geq150$ & 5253.7 & 168.0 & 36.67 & 0 & 0 & 0 \\
        
        $M_{b_1b_2\ell_1}\geq260$ & 5142.8 & 131.2 & 23.36 & 0 & 0 & 0 \\
        
        $\cancel{H}_T\geq110$ & 5086.4 & 121.6 & 23.36 & 0 & 0 & 0 \\
        \bottomrule
\end{tabular}
\end{minipage}
\caption{Cutflow results for Signal~4 at the LHeC for two benchmark points: BP--1 (\(M_{H^\pm}=300~\mathrm{GeV}\)) and BP--2 (\(M_{H^\pm}=500~\mathrm{GeV}\)).}
\label{tab:sig4LHec:combained}
\end{table}

\begin{table}[t]
\centering
\begin{minipage}{0.47\linewidth} 
\centering
\begin{tabular}{!{\vrule}c!{\vrule}c!{\vrule}c!{\vrule}c!{\vrule}c!{\vrule}c!{\vrule}c!{\vrule}}
\toprule
\multicolumn{7}{!{\vrule}c!{\vrule}}{\textbf{(a) BP--1: \(M_{H^\pm}=300~\mathrm{GeV}\)}} \\
\midrule
\textbf{Cuts} & \textbf{S} & \textbf{BG1} & \textbf{BG2} & \textbf{BG3} & \textbf{BG4} & \textbf{BG5} \\
\midrule
        -- & 100000 & 200000 & 100000 & 100000 & 100000 & 100000 \\
        
        \( N_b \geq 2 \) & 16288 & 20449 & 7949.5 & 16663 & 23180 & 8248.6 \\
        
        \( N_j \geq 1 \) & 13477 & 16061 & 5908.6 & 16480 & 22956 & 8246.6 \\
        
        \( N_\ell \geq 1 \) & 6762.6 & 7918.3 & 2906.8 & 3.01 & 6.65 & 3.0 \\
        
        $M_{b_1b_2\ell_1}\geq160$ & 5506.8 & 5264.9 & 1915.7 & 0.603 & 4.4 & 2.0 \\
        
        $\cancel{H}_T\geq60$ & 4621.9 & 3620.0 & 1309.3 & 0 & 2.21 & 0.998 \\
        \bottomrule
\end{tabular}
\end{minipage}%
\hfill
\begin{minipage}{0.47\linewidth} 
\centering
\begin{tabular}{!{\vrule}c!{\vrule}c!{\vrule}c!{\vrule}c!{\vrule}c!{\vrule}c!{\vrule}c!{\vrule}}
\toprule
\multicolumn{7}{!{\vrule}c!{\vrule}}{\textbf{(b) BP--2: \(M_{H^\pm}=500~\mathrm{GeV}\)}} \\
\midrule
\textbf{Cuts} & \textbf{S} & \textbf{BG1} & \textbf{BG2} & \textbf{BG3} & \textbf{BG4} & \textbf{BG5} \\
\midrule
        -- & 100000 & 200000 & 100000 & 100000 & 100000 & 100000 \\
        
        \( N_b \geq 2 \) & 15358 & 20049 & 7949.5 & 16663 & 23180 & 8248.6 \\
        
        \( N_j \geq 1 \) & 13461 & 16061 & 5908.6 & 16480 & 22956 & 8246.6 \\
        
        \( N_\ell \geq 1 \) & 4869.1 & 7918.3 & 2906.8 & 3.01 & 6.65 & 3.0 \\
        
        $P_T(b_1)\geq100$ & 4070.5 & 2651.6 & 984.3 & 1.2 & 3.32 & 0.997 \\
        
        $P_T(\ell_1)\geq70$ & 3028.9 & 658.9 & 231.4 & 0 & 0 & 0 \\
        
        $\Delta R(b_1,b_2)\leq1.8$ & 2883.9 & 198.4 & 65.38 & 0 & 0 & 0 \\
        
        $M_{b_1\ell_1}\geq150$ & 2881.8 & 178.7 & 58.20 & 0 & 0 & 0 \\
        
        $M_{b_1b_2\ell_1}\geq260$ & 2818.5 & 130.9 & 40.38 & 0 & 0 & 0 \\
        
        $\cancel{H}_T\geq110$ & 2761.3 & 116.7 & 35.35 & 0 & 0 & 0 \\
        \bottomrule
\end{tabular}
\end{minipage}
\caption{Cutflow results for Signal~4 at the FCC-eh for two benchmark points: BP--1 (\(M_{H^\pm}=300~\mathrm{GeV}\)) and BP--2 (\(M_{H^\pm}=500~\mathrm{GeV}\)).}
\label{tab:sig4FCCeh:combained}
\end{table}

We emphasize that the preceding analysis primarily served as an exercise to demonstrate the suppression of SM backgrounds for the relevant processes. The signal cross sections considered so far were fiducial and not tied to a specific model. Before concluding this section, it is instructive to determine the minimum signal cross section required for discovery or exclusion within a particular model, for each of the four signals and both benchmark points, under the LHeC and FCC-eh collider configurations. For this purpose, we employ the statistical measures proposed in~\cite{Cowan:2010js}.
In this formalism, two significance measures are commonly used: 
$\mathcal{Z}_D$, referred to as the \emph{discovery significance}, quantifies the probability 
of observing a signal-like excess over the background under the signal-plus-background hypothesis. 
Conversely, $\mathcal{Z}_E$, the \emph{exclusion significance}, measures the metric at which to 
exclude the presence of a signal when only the background hypothesis is true. They are given by
\begin{equation}
\mathcal{Z}_D = \sqrt{ 2 \left[ (S + B)\,\log\!\left(1 + \frac{S}{B}\right) - S \right] }\,,
\end{equation}
\begin{equation}
\mathcal{Z}_E = \sqrt{ -2 \left( B\,\log\!\left( 1 + \frac{S}{B} \right) - S \right) }\,,
\end{equation}

where $S = \sigma_s \mathcal{L}$ and $B = \sigma_B \mathcal{L}$ denote the total number of signal and background events that survive the selection cuts. In \autoref{tab:scross1} for LHeC and \autoref{tab:scross2} for FCC-eh, we present the required signal cross sections for both discovery and exclusion scenarios. It is worth noting, however, that the interpretation of a $5\sigma$ discovery potential depends strongly on the underlying model—particularly the coupling strengths and decay patterns of the charged Higgs—and thus can only be fully addressed within a model-specific framework. We will turn our attention to this aspect in the subsequent discussion.

\begin{table}[h]
    \centering
    \renewcommand{\arraystretch}{1.2}
    \begin{tabular}{ !{\vrule}c !{\vrule} c !{\vrule} c !{\vrule} c !{\vrule} c !{\vrule} c !{\vrule} c !{\vrule} c !{\vrule} c !{\vrule} c !{\vrule} c !{\vrule} c !{\vrule} } 
        \toprule
        \textbf{Signal} & \textbf{BP} & \textbf{Background (fb)} &
        \multicolumn{3}{c!{\vrule}}{\(\mathcal{Z}_E \geq 1.96\)} & 
        \multicolumn{3}{c!{\vrule}}{\(\mathcal{Z}_D \geq 3\sigma\)} & 
        \multicolumn{3}{c!{\vrule}}{\(\mathcal{Z}_D \geq 5\sigma\)} \\
        \cmidrule{4-12}
        & & & 
        \(\mathcal{L} = 250\) & \(\mathcal{L} = 500 \) & \(\mathcal{L} = 1000\) &
        \(\mathcal{L} = 250\) & \(\mathcal{L} = 500\) & \(\mathcal{L} = 1000\) &
        \(\mathcal{L} = 250\) & \(\mathcal{L} = 500\) & \(\mathcal{L} = 1000\) \\
        \midrule
        Signal 1 & BP1 & 22.931 & 0.5987 & 0.4223 & 0.2980 & 0.9145 & 0.6454 & 0.4557 & 1.5308 & 1.0790 & 0.7613 \\
        
         & BP2 & 3.819 & 0.2473 & 0.1738 & 0.1224 & 0.3767 & 0.2651 & 0.1868 & 0.6344 & 0.4452 & 0.3131 \\
        \midrule
        Signal 2 & BP1 & 21.152 & 0.5752 & 0.4056 & 0.2863 & 0.8786 & 0.6200 & 0.4378 & 1.4709 & 1.0366 & 0.7313 \\
        
         & BP2 & 0.244 & 0.0664 & 0.0458 & 0.0319 & 0.0995 & 0.0692 & 0.0483 & 0.1721 & 0.1185 & 0.0821 \\
        \midrule
        Signal 3 & BP1 & 3.100 & 0.2234 & 0.1569 & 0.1104 & 0.3400 & 0.2392 & 0.1685 & 0.5732 & 0.4019 & 0.2825 \\
        
         & BP2 & 0.297 & 0.0727 & 0.0503 & 0.0350 & 0.1092 & 0.0760 & 0.0531 & 0.1883 & 0.1299 & 0.0902 \\
        \midrule
        Signal 4 & BP1 & 2.553 & 0.2032 & 0.1426 & 0.1003 & 0.3091 & 0.2173 & 0.1530 & 0.5216 & 0.3655 & 0.2567 \\
        
         & BP2 & 0.058 & 0.0351 & 0.0237 & 0.0162 & 0.0513 & 0.0351 & 0.0243 & 0.0914 & 0.0616 & 0.0420 \\
        \bottomrule
    \end{tabular}
    \caption{Required signal cross sections (in fb) at different luminosities~($fb^{-1}$) to achieve exclusion (\(\mathcal{Z}_E \geq 1.96\)) and discovery (\(\mathcal{Z}_D \geq 3\sigma,\ 5\sigma\)) significance for given background yields (in fb) at LHeC collider setup}
    \label{tab:scross1}
\end{table}

\begin{table}[h]
    \centering
    \renewcommand{\arraystretch}{1.2}
    \begin{tabular}{ !{\vrule}c !{\vrule} c !{\vrule} c !{\vrule} c !{\vrule} c !{\vrule} c !{\vrule} c !{\vrule} c !{\vrule} c !{\vrule} c !{\vrule} c !{\vrule} c !{\vrule} } 
        \toprule
        \textbf{Signal} & \textbf{BP} & \textbf{Background (fb)} &
        \multicolumn{3}{c!{\vrule}}{\(\mathcal{Z}_E \geq 1.96\)} & 
        \multicolumn{3}{c!{\vrule}}{\(\mathcal{Z}_D \geq 3\sigma\)} & 
        \multicolumn{3}{c!{\vrule}}{\(\mathcal{Z}_D \geq 5\sigma\)} \\
        \cmidrule{4-12}
        & & & 
        \(\mathcal{L} = 500\) & \(\mathcal{L} = 1000\) & \(\mathcal{L} = 2000\) &
        \(\mathcal{L} = 500\) & \(\mathcal{L} = 1000\) & \(\mathcal{L} = 2000\) &
        \(\mathcal{L} = 500\) & \(\mathcal{L} = 1000\) & \(\mathcal{L} = 2000\) \\
        \midrule
        Signal 1 & BP1 & 290.007 & 1.4952 & 1.0567 & 0.7469 & 2.2877 & 1.6170 & 1.1431 & 3.8162 & 2.6967 & 1.9060 \\
        
         & BP2 & 48.268 & 0.6115 & 0.4318 & 0.3051 & 0.9351 & 0.6605 & 0.4668 & 1.5618 & 1.1026 & 0.7788 \\
        \midrule
        Signal 2 & BP1 & 290.001 & 1.4952 & 1.0567 & 0.7469 & 2.2877 & 1.6170 & 1.1431 & 3.8162 & 2.6967 & 1.9060 \\
        
         & BP2 & 11.913 & 0.3051 & 0.2152 & 0.1519 & 0.4660 & 0.3289 & 0.2322 & 0.7800 & 0.5498 & 0.3879 \\
        \midrule
        Signal 3 & BP1 & 35.826 & 0.5272 & 0.3722 & 0.2629 & 0.8060 & 0.5693 & 0.4022 & 1.3467 & 0.9505 & 0.6712 \\
        
         & BP2 & 5.286 & 0.2040 & 0.1437 & 0.1014 & 0.3114 & 0.2196 & 0.1549 & 0.5223 & 0.3676 & 0.2591 \\
        \midrule
        Signal 4 & BP1 & 30.420 & 0.4860 & 0.3431 & 0.2423 & 0.7429 & 0.5247 & 0.3707 & 1.2415 & 0.8762 & 0.6187 \\
        
         & BP2 & 0.894 & 0.0854 & 0.0598 & 0.0420 & 0.1298 & 0.0911 & 0.0641 & 0.2196 & 0.1536 & 0.1077 \\
        \bottomrule
    \end{tabular}
    \caption{Required signal cross sections (in fb) at different luminosities to achieve exclusion (\(\mathcal{Z}_E \geq 1.96\)) and discovery (\(\mathcal{Z}_D \geq 3\sigma,\ 5\sigma\)) significance for given background yields (in fb) in FCC-eh collider setup}
    \label{tab:scross2}
\end{table}

\section{Phenomenological Implications}
\label{tab:PI}

We now turn to the phenomenological implications of our analysis in the context of a BSM scenario featuring an extended gauge symmetry  
\(
SU(2)_0 \times SU(2)_1 \times U(1)_2
\), as proposed in Ref.~\cite{Coleppa:2020set}.  
This framework introduces an enlarged scalar sector that naturally accommodates a charged Higgs boson, along with additional heavy gauge bosons.  
In this section, we focus on the collider reach of the previously discussed signals within the parameter space of this specific model.  
For clarity, we summarize only the key ingredients relevant to our study; a detailed and comprehensive description of the model can be found in Ref.~\cite{Coleppa:2020set}.

The extended gauge symmetry leads to new heavy vector bosons and additional charged as well as neutral Higgs states.  
Electroweak symmetry breaking (EWSB) in this setup is realized through two Higgs doublets, \(\Phi_{1,2}\), supplemented by a nonlinear sigma model field \(\Sigma\).  The $\Sigma$ field transforms under the two $SU(2)$ gauge groups as $\Sigma\to G_0^{\dagger}\Sigma G_1$, and the two additional scalar doublets in the model $\Phi_{1,2}$ transform as ($\textbf{2},\textbf{1},\frac{1}{2}$) and ($\textbf{1},\textbf{2},\frac{1}{2}$) respectively under the full symmetry group of the model.
The vacuum expectation values (vevs) of these fields are denoted as \(F\) for \(\Sigma\) and \(\Phi_2\), while \(f\) corresponds to the VEV of \(\Phi_1\).  
These are conveniently expressed in terms of the electroweak scale \(v\) as  
\[
F = \sqrt{2}\, v \cos\beta, \qquad f = v \sin\beta\,.
\]

The explicit forms of the scalar fields are given by\footnote{Here, we identify $f_1 = f$ and $f_2 = F$.}  
\[
\Sigma = \exp\!\left( \frac{i \Pi_\Sigma^a \sigma^a}{F} \right), \quad 
\Phi_b = \begin{pmatrix} 
\dfrac{f_b + H_b + i \Pi_b^0}{\sqrt{2}} \\[4pt] 
i \Pi_b^- 
\end{pmatrix}, \quad b = 1,2\,.
\]

The gauge couplings corresponding to the groups \(SU(2)_0\), \(SU(2)_1\), and \(U(1)_2\) are denoted by \(g_0\), \(g_1\), and \(g_2\), respectively. They are related to the electroweak parameters through  
\[
g = \frac{e}{\sin\theta \cos\phi}, \qquad 
g_1 = \frac{e}{\sin\theta \sin\phi}, \qquad 
g_2 = \frac{e}{\cos\theta}\,.
\]

The most general gauge-invariant scalar potential can be written as  
\begin{align}
\begin{split}
V(\Phi_{1},\Phi_{2},\Sigma) &= 
\lambda_{1}\!\left[\Phi^{\dagger}_{1}\Phi_{1} - \frac{f^{2}}{2}\right]^{2} 
+ \lambda_{2}\!\left[\Phi^{\dagger}_{2}\Phi_{2} - \frac{F^{2}}{2}\right]^{2} 
+ \lambda_{3}\!\left[\Phi^{\dagger}_{1}\Phi_{1} + \Phi^{\dagger}_{2}\Phi_{2} - \frac{f^{2} + F^{2}}{2}\right]^{2} \\
&\quad + \lambda_{4}\!\left[(\Phi^{\dagger}_{1}\Phi_{1})(\Phi^{\dagger}_{2}\Phi_{2}) - (\Phi^{\dagger}_{1}\Sigma\Phi_{2})(\Phi^{\dagger}_{2}\Sigma^{\dagger}\Phi_{1})\right] 
+ \lambda_{5}\!\left[\mathrm{Re}(\Phi^{\dagger}_{1}\Sigma\Phi_{2}) - \frac{f F}{2}\right]^{2} \\
&\quad + \lambda_{6}\,\mathrm{Im}\!\left[\Phi^{\dagger}_{1}\Sigma\Phi_{2}\right]^{2}\,.
\end{split}
\label{eqn:L2}
\end{align}

After EWSB, the scalar spectrum contains two CP-even Higgs bosons (\(h\), \(H\)), a pseudoscalar \(A\), and a pair of charged Higgs bosons \(H^{\pm}\).  
Notably, the mass of the charged Higgs originates entirely from the \(\lambda_{4}\) term in Eqn.~\ref{eqn:L2}, whereas the \(\lambda_{5}\) and \(\lambda_{6}\) terms generate only interaction terms without quadratic mass contributions for the pion fields.
The mass matrix for the charged scalars takes the form  
\begin{equation}
M^{2}_{\pi^{\pm}} = \frac{\lambda_{4}}{2}
\begin{bmatrix}
f^{2} & -fF & f^{2} \\
-fF   & F^{2} & -fF \\
f^{2} & -fF & f^{2}
\end{bmatrix}.
\end{equation}
The eigenstates corresponding to the Goldstone modes, \(G_{1}^{\pm}\) and \(G_{2}^{\pm}\), which are absorbed as the longitudinal components of \(W^{\pm}_{\mu}\) and \(W^{'\pm}_{\mu}\), are given by  
\begin{align}
\begin{split}
G_{1}^{\pm} &= -\frac{1}{\sqrt{2}}\Pi_{\Sigma}^{\pm} + \frac{1}{\sqrt{2}}\Pi_{2}^{\pm},\,\, \textrm{and}\\
G_{2}^{\pm} &= \frac{F}{2v}\Pi_{\Sigma}^{\pm} + \frac{f}{v}\Pi_{1}^{\pm} + \frac{F}{2v}\Pi_{2}^{\pm}\,.
\label{eqn:goldstone}
\end{split}
\end{align}

The orthogonal combination corresponds to the physical charged Higgs boson,  
\begin{equation}
H^{\pm} = \frac{f}{\sqrt{2}v}\,\Pi_{\Sigma}^{\pm} - \frac{F}{\sqrt{2}v}\,\Pi_{1}^{\pm} + \frac{f}{\sqrt{2}v}\,\Pi_{2}^{\pm},  
\label{eqn:chargedHiggs}
\end{equation}
with mass  
\[
M^{2}_{H^{\pm}} = \frac{\lambda_{4}}{2}\left(2f^{2} + F^{2}\right)\,.
\]

The fermionic sector is constructed such that it mimics the Type-I 2HDM Higgs–fermion couplings at leading order, ensuring that SM fermions acquire their masses primarily through the \(\Phi_{1}\) doublet.  In Table~\ref{tab:Hc_Coupling}, we summarize the relevant charged Higgs boson couplings in terms of the parameter \(x = m_W/m_{W'}\).  
The lighter CP-even scalar \(h\) is identified with the observed 125~GeV Higgs boson, while the heavier neutral Higgs states are assumed to lie above the charged Higgs in mass.  
As a result, decays such as \(H^{\pm} \to VH/VA\) (where \(V = W\) or \(W'\)) are kinematically forbidden. The mixing angle between the two CP-even states is denoted by \(\alpha\), following the usual 2HDM convention.  
The coupling \(\xi^{H^{\pm}}_{ff'}\) remains proportional to the ratio of the two VEVs, similar to a Type-I 2HDM. An interesting feature of this model is the presence of a non-zero \(\xi^{H^{\pm}}_{W^{\mp}Z}\) coupling at tree level, which is absent in models where symmetry breaking is induced solely by scalar singlets or doublets (see~\autoref{app} for a detailed discussion of this feature).

\begin{table}[h]
    \centering
    \begin{tabular}{ !{\vrule}c !{\vrule} c !{\vrule} } 
        \toprule
        \textbf{Coupling} & \textbf{Expression} \\ 
        \midrule
        $\xi^{H^{\pm}}_{W^{'\mp}Z}$ & $\dfrac{\sin\beta}{2}\!\left(1 + \dfrac{x^{2}}{4}\right)$ \\
        \midrule
        $\xi^{H^{\pm}}_{W^{\mp}Z}$ & $\dfrac{x^{2}\cos\beta\sin\beta}{16\sin^{2}\theta_{w}\cos^{2}\theta_{w}}$ \\
        \midrule
        $\xi^{H^{\pm}}_{W^{\mp}Z^{'}}$ & $\dfrac{\sin\beta\cos\theta_{w}}{2}\!\left(1 + \dfrac{x^{2}}{8}\right)$ \\
        \midrule
        $\xi^{H^{\pm}}_{ff^{'}}$ & $\cot\beta\!\left(1 - \dfrac{x^{2}}{4}\right)$ \\
        \midrule
        $\xi^{H^{\pm}}_{W^{\mp}h}$ & $\dfrac{1}{4}\!\Big[(4\sin\alpha\cos\beta + \sqrt{2}\sin\alpha\sin\beta) + \dfrac{x^{2}}{32}\,(8\sin\alpha\cos\beta - \sqrt{2}\sin\alpha\sin\beta) \Big]$ \\
        \bottomrule
    \end{tabular}
    \caption{Charged Higgs boson couplings relevant for the phenomenology of the model discussed in Ref.~\cite{Coleppa:2020set}.}
    \label{tab:Hc_Coupling}
\end{table}

\begin{figure}[h!]
    \centering
    \includegraphics[width=0.45\linewidth]{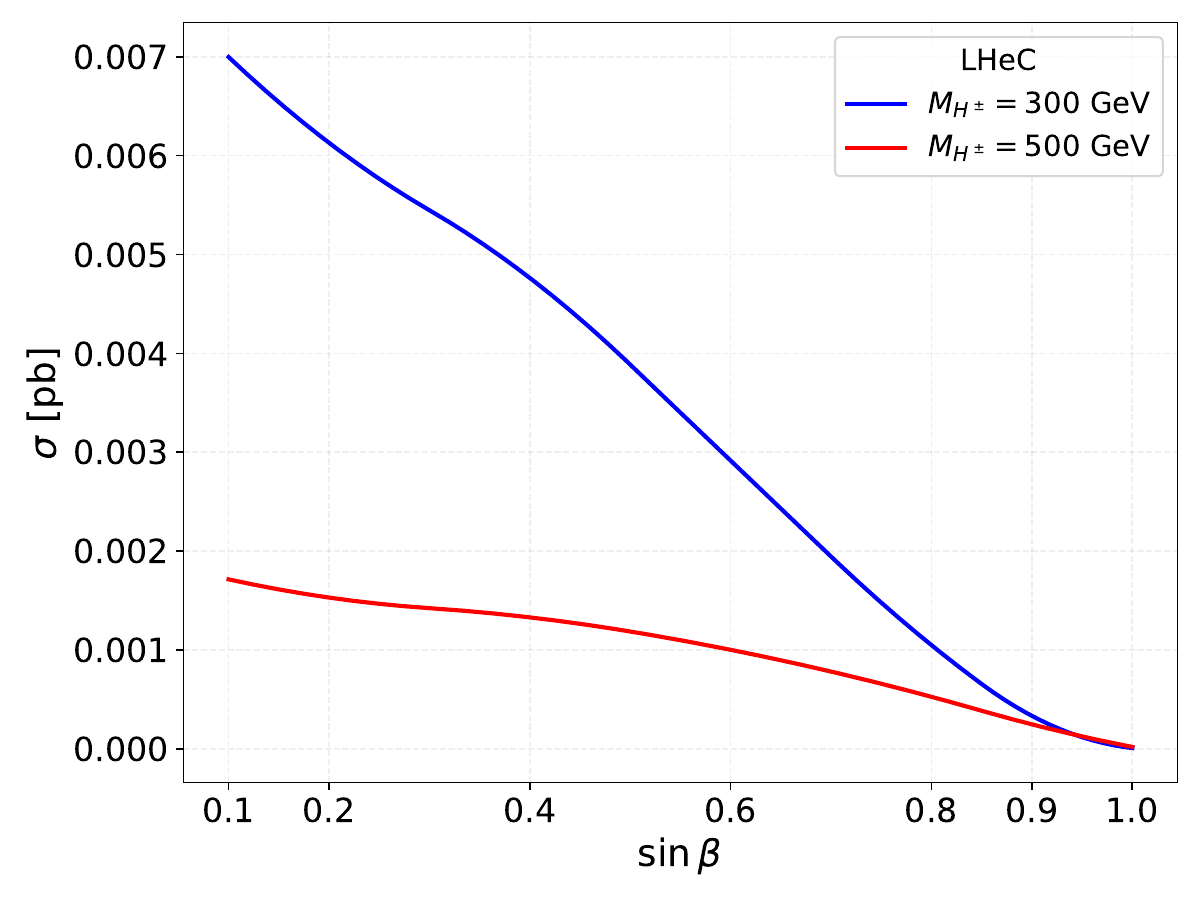}
    \includegraphics[width=0.45\linewidth]{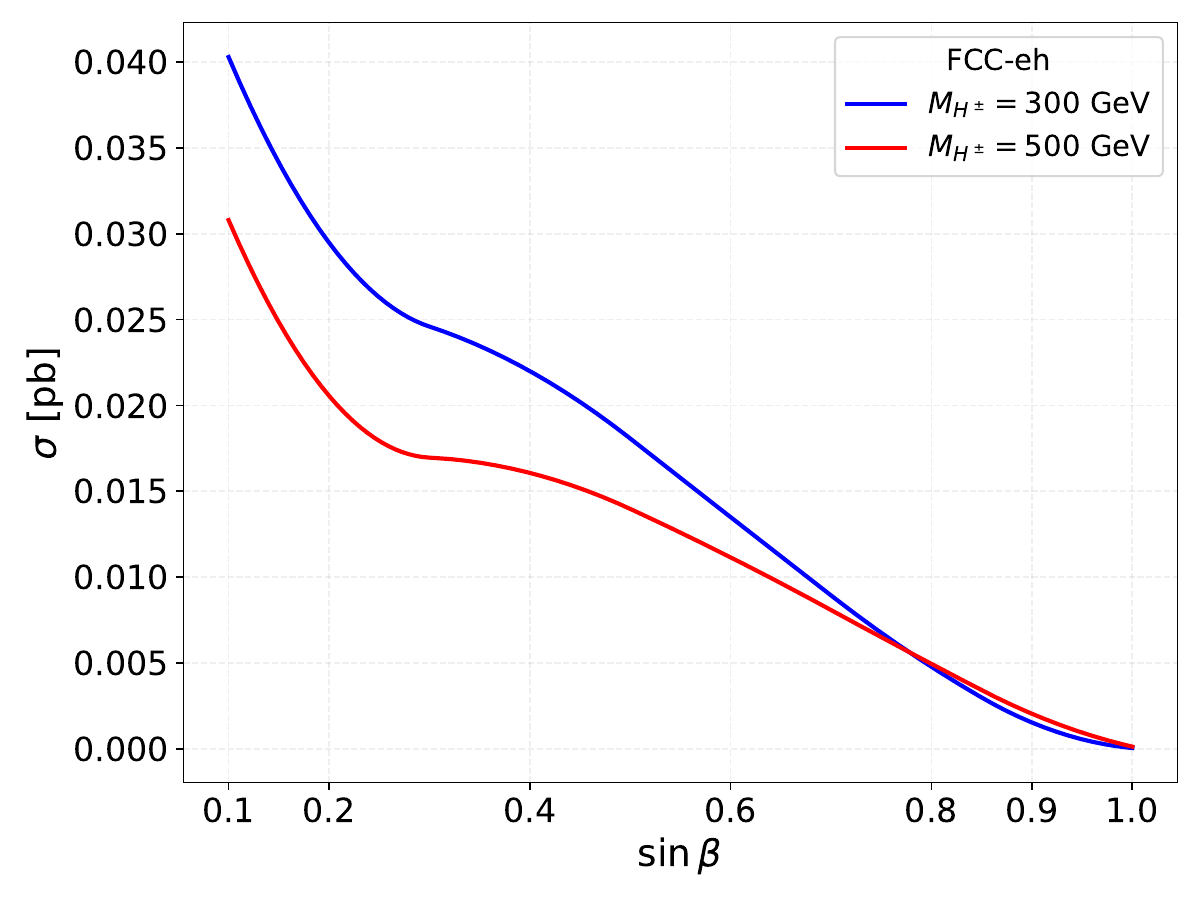}
    \caption{Charged Higgs production cross section for the process $e^- p \to \nu_e\, H^-\, j$ as a function of $\sin\beta$. \textbf{Left:} LHeC setup; \textbf{Right:} FCC-eh setup.}
    \label{fig:Prod}
\end{figure}

We now turn to the phenomenology of the charged Higgs boson within this model. In \autoref{fig:Prod}, we present the production cross section of the charged Higgs in the process $e^- p \to \nu_e\, H^-\, j$ for two benchmark mass scenarios, $M_{H^-} = 450~\text{GeV}$ and $550~\text{GeV}$, as a function of $\sin\beta$, which parameterizes the ratio of the vacuum expectation values (vevs). Results are shown for both the LHeC and FCC-eh, evaluated at their respective center-of-mass energies.  

\begin{figure}[h!]
    \centering
    \includegraphics[width=0.45\linewidth]{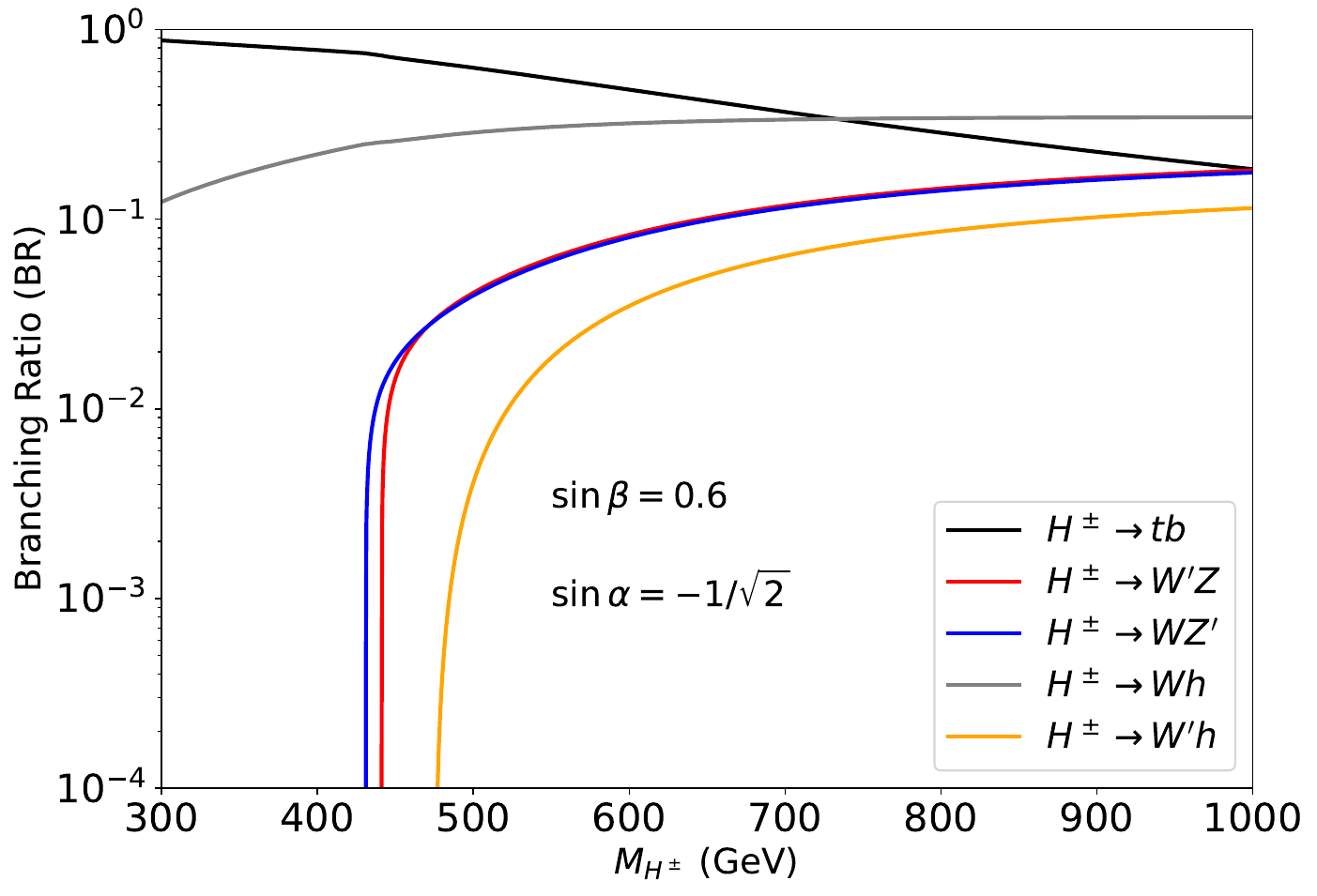}
    \includegraphics[width=0.45\linewidth]{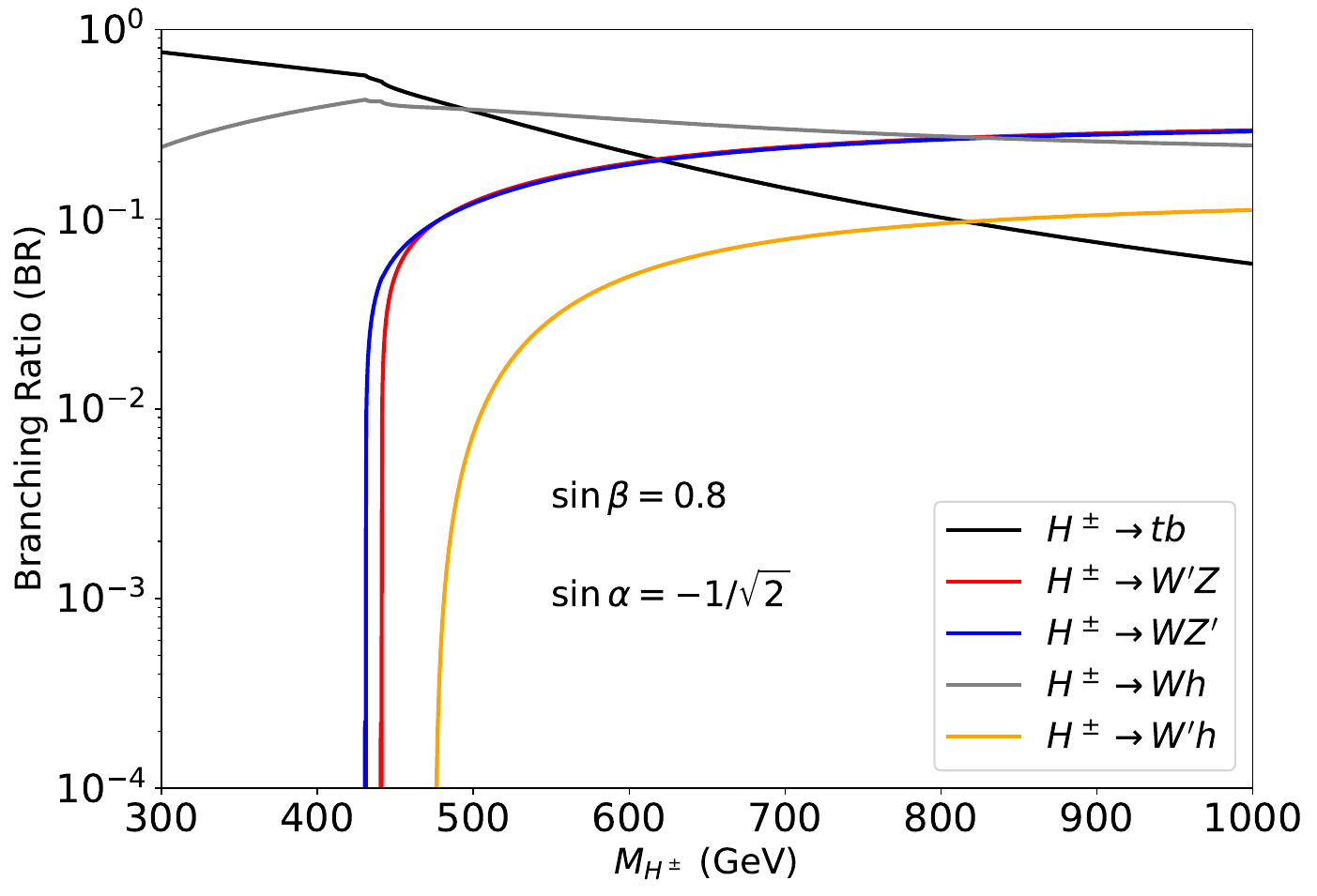}
    \caption{Branching ratios of the charged Higgs boson into various decay modes for two representative values of $\sin\beta$: \textbf{Left:} $\sin\beta = 0.6$; \textbf{Right:} $\sin\beta = 0.8$.}
    \label{fig:BR}
\end{figure}

In \autoref{fig:BR}, we illustrate the branching ratios of $H^\pm$ into the various kinematically allowed decay channels. The mixing angle between the two CP-even neutral Higgs bosons, $\alpha$, is chosen such that $\sin\alpha = -1/\sqrt{2}$. This specific choice is compatible with current Higgs data, provided $\sin\beta > 0.4$. For a more detailed discussion of the theoretical and experimental constraints justifying this parameter selection, we refer the reader to Ref.~\cite{Coleppa:2020set}.  

\begin{figure}[h!]
    \centering
    \includegraphics[width=0.40\linewidth]{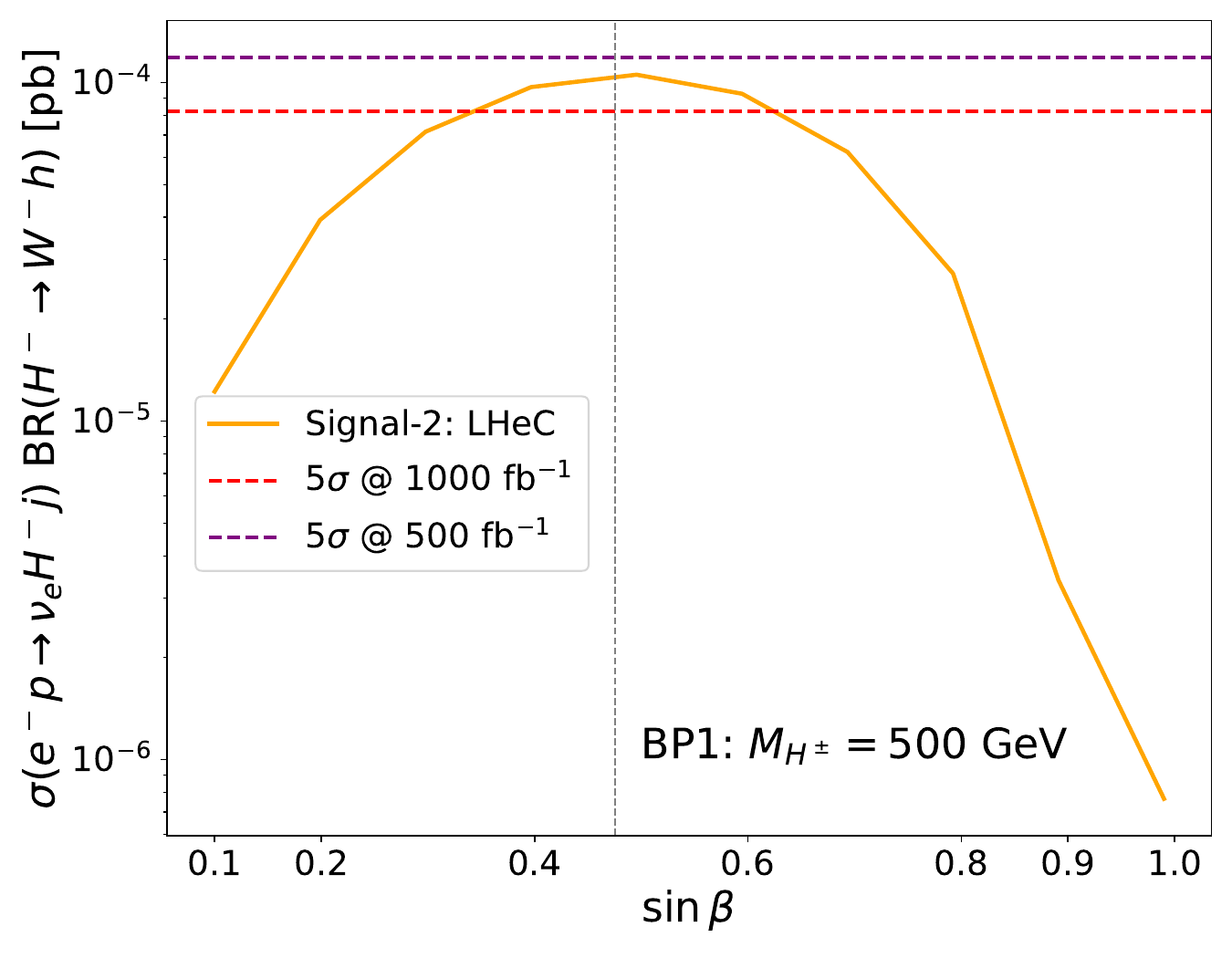}
    \includegraphics[width=0.40\linewidth]{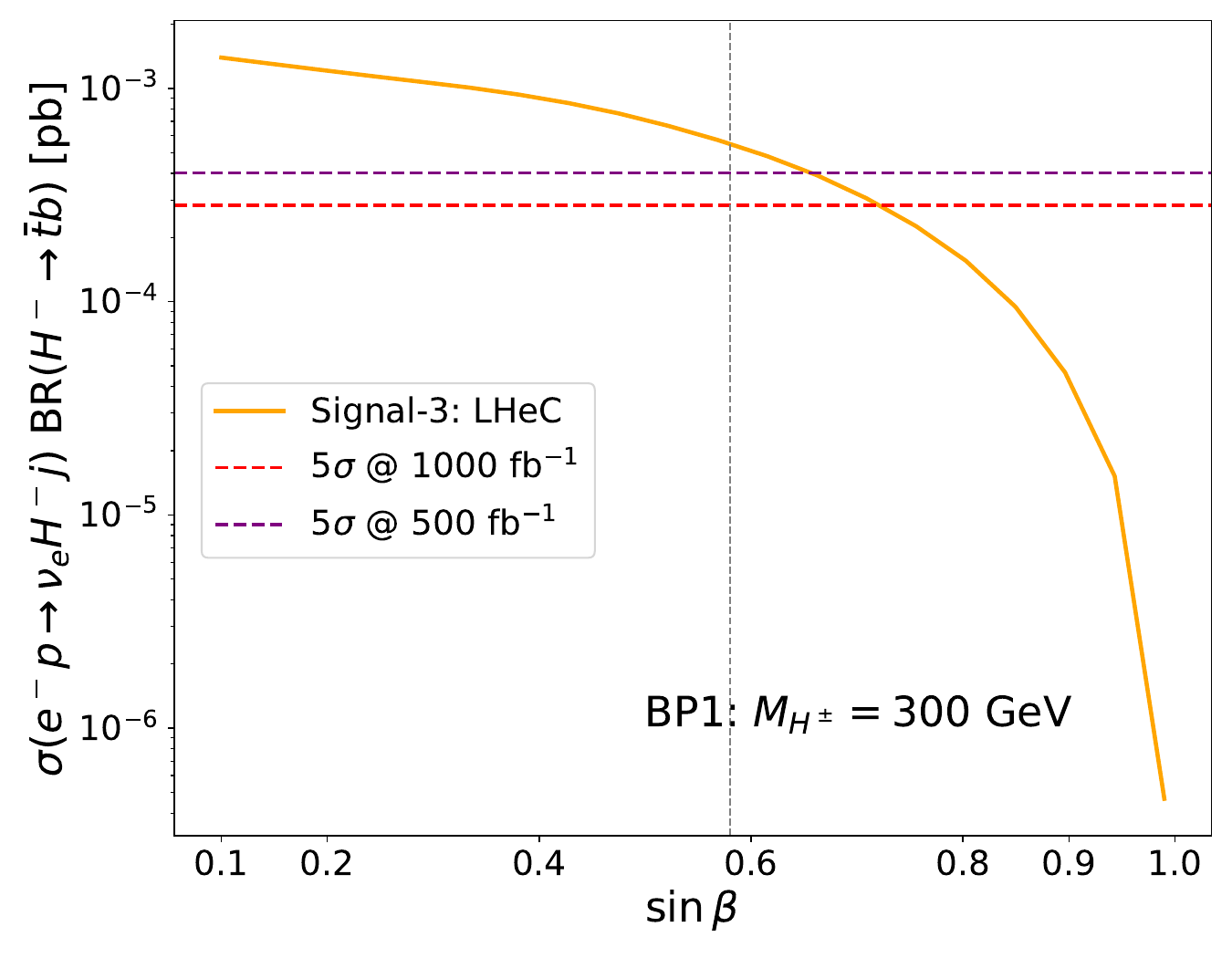}
    \includegraphics[width=0.40\linewidth]{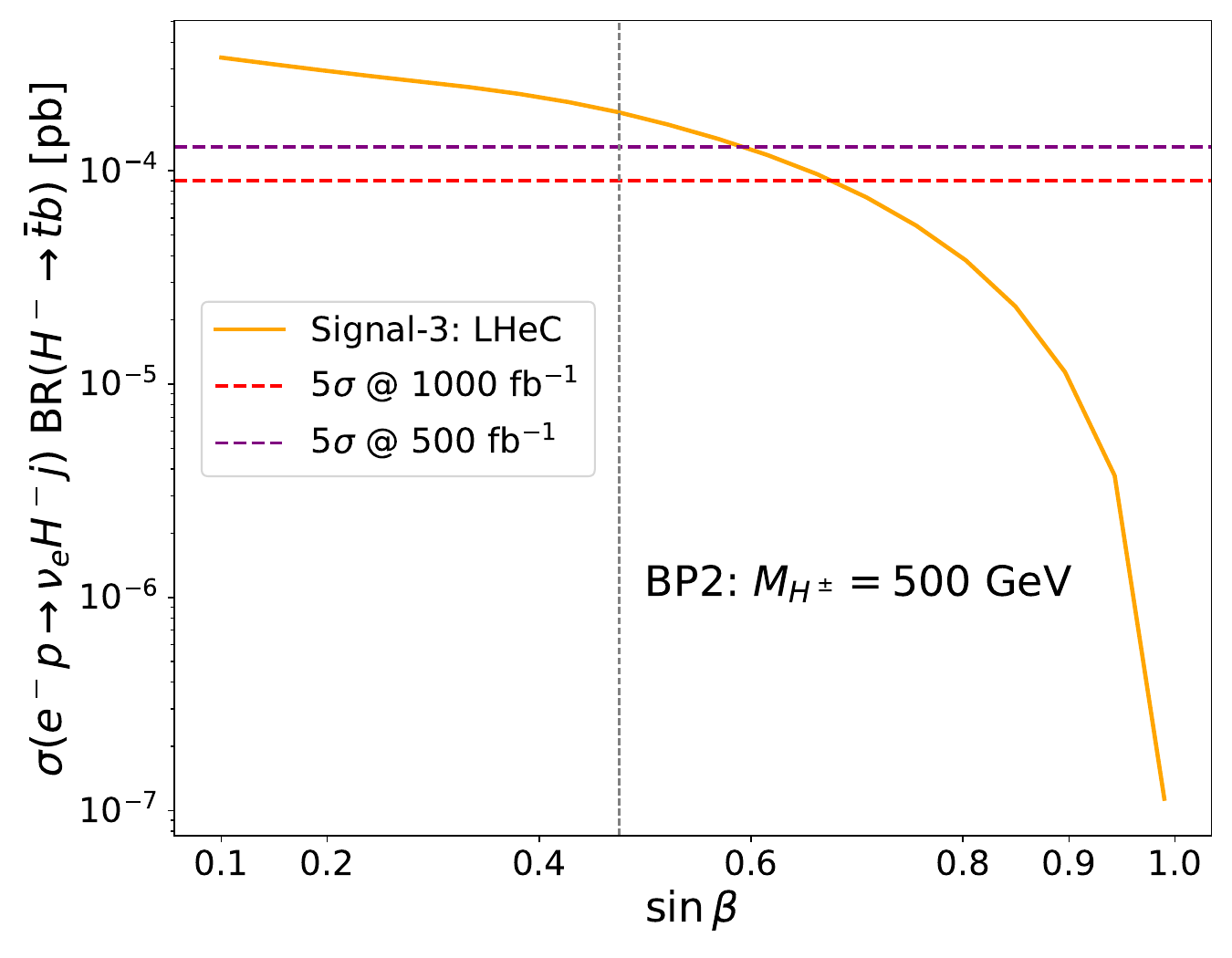}
    \includegraphics[width=0.40\linewidth]{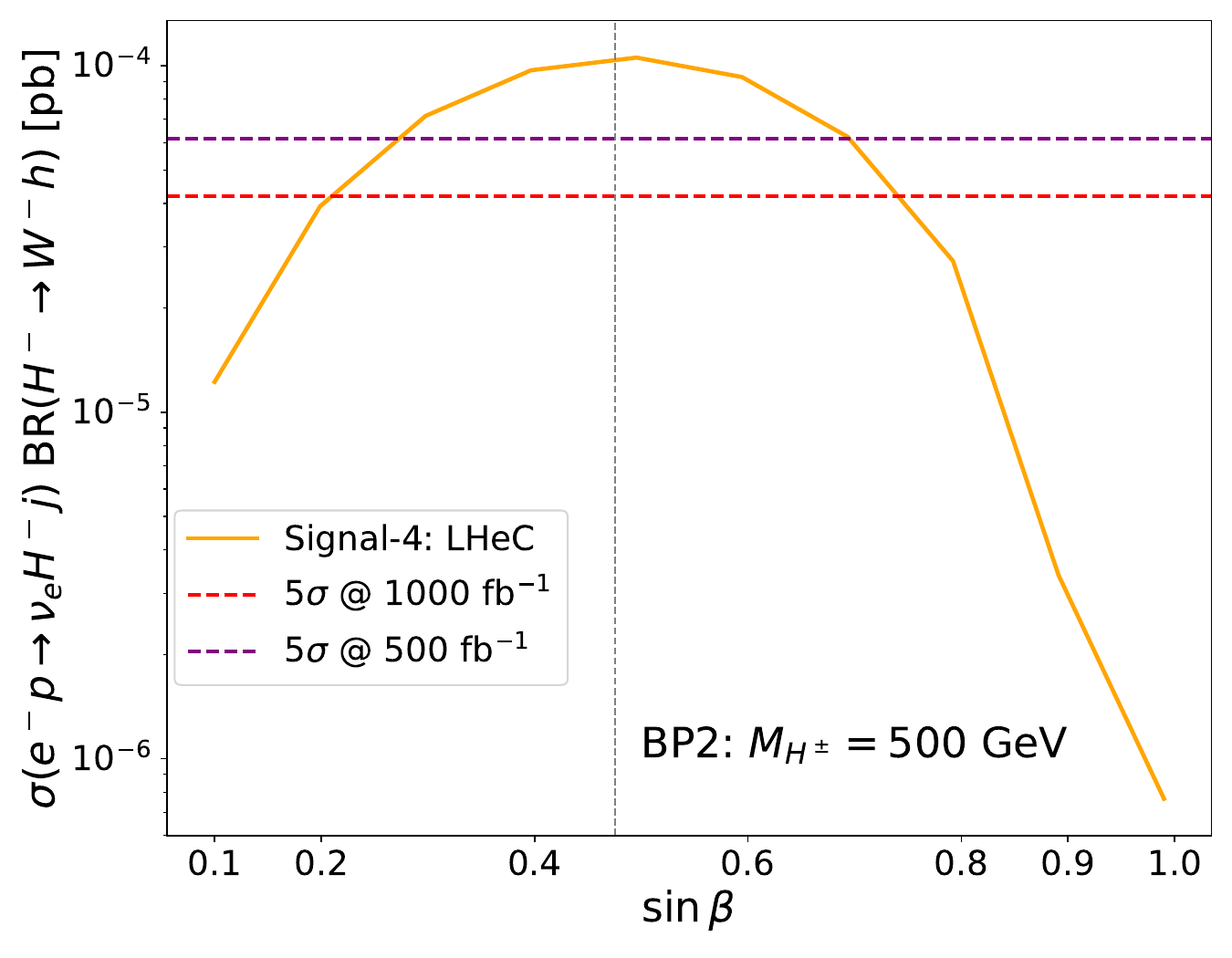}
    \caption{Reach plots illustrating the accessible parameter space at the LHeC for signal~3 and signal~4 for both benchmark points, shown as a function of $\sin{\beta}$ for $\sin{\alpha} = \tfrac{-1}{\sqrt{2}}$. The region to the left of the gray dashed vertical line is excluded by constraints from $b \to s\gamma$.}
    \label{fig:RP-LHeC}
\end{figure}

\begin{figure}[H]
    \centering
    \includegraphics[width=0.32\linewidth]{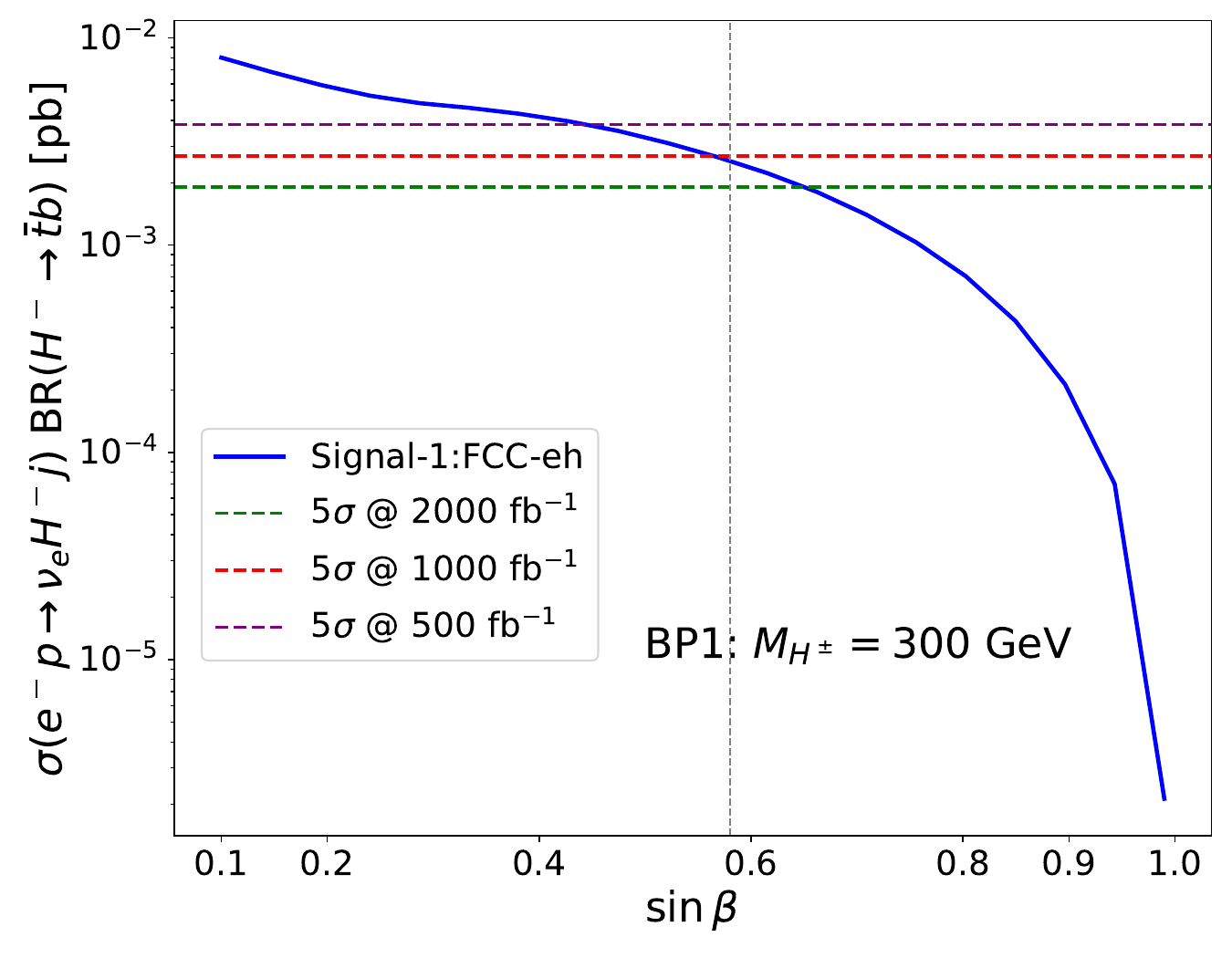}
    \includegraphics[width=0.32\linewidth]{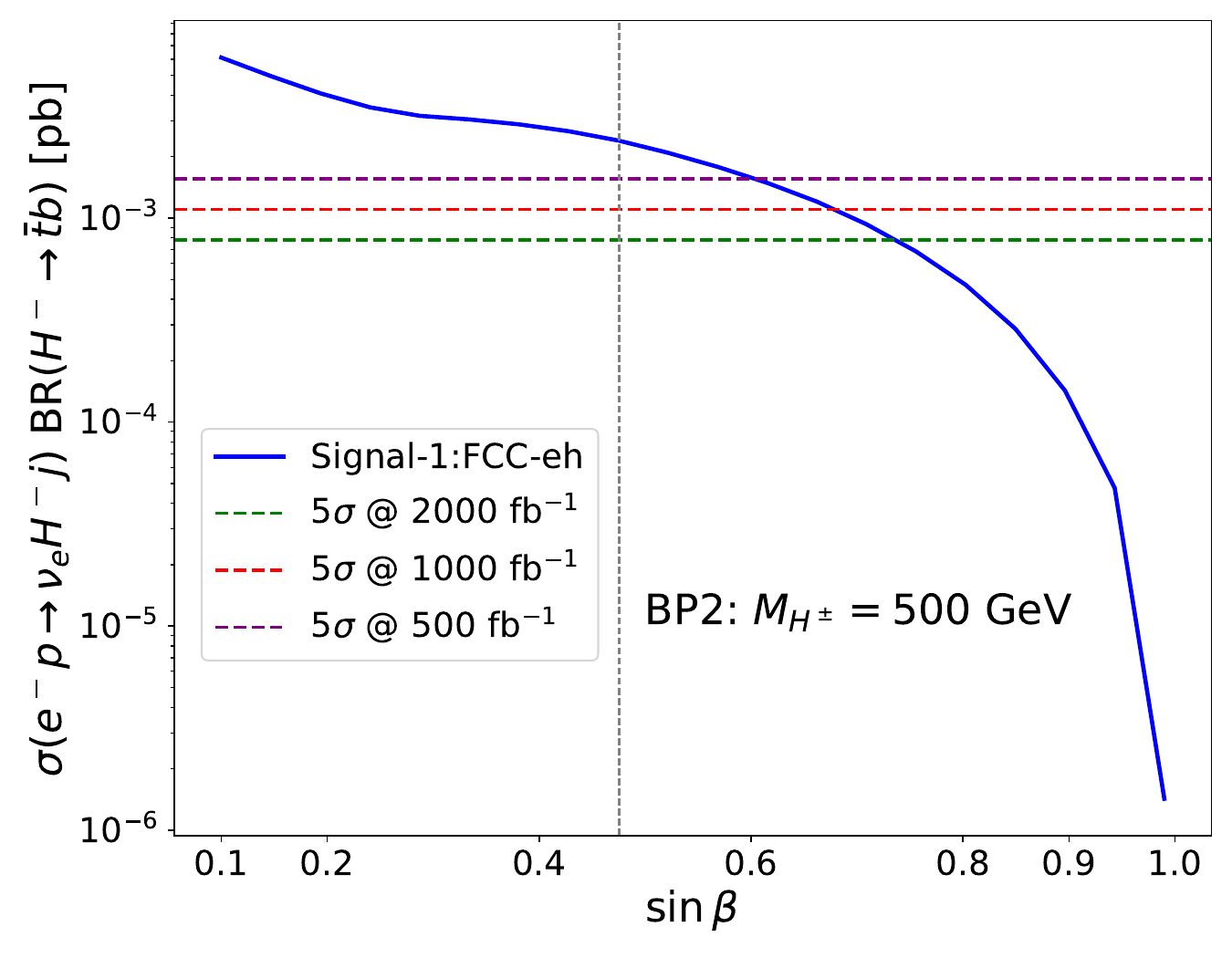}
    \includegraphics[width=0.32\linewidth]{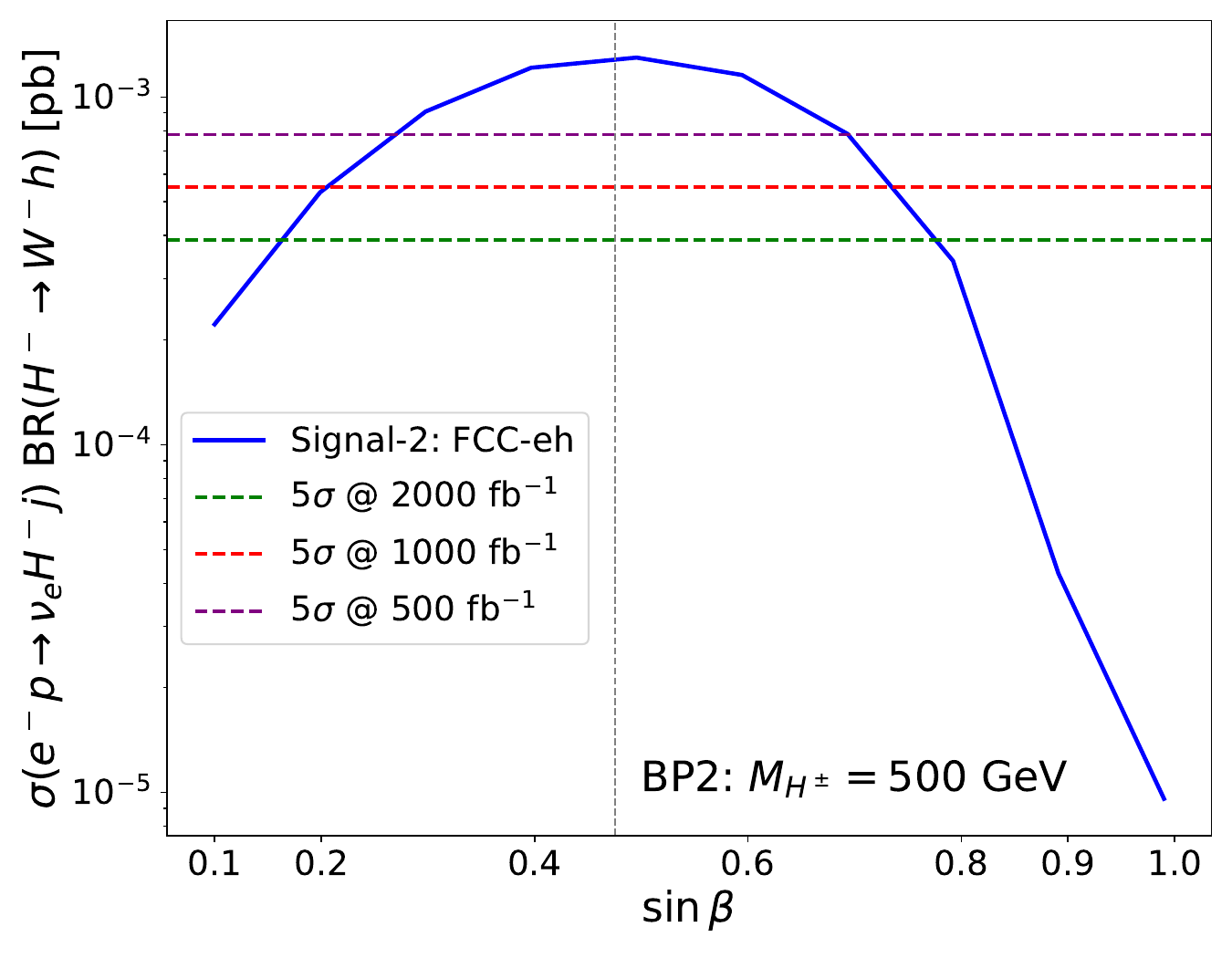}
    \includegraphics[width=0.32\linewidth]{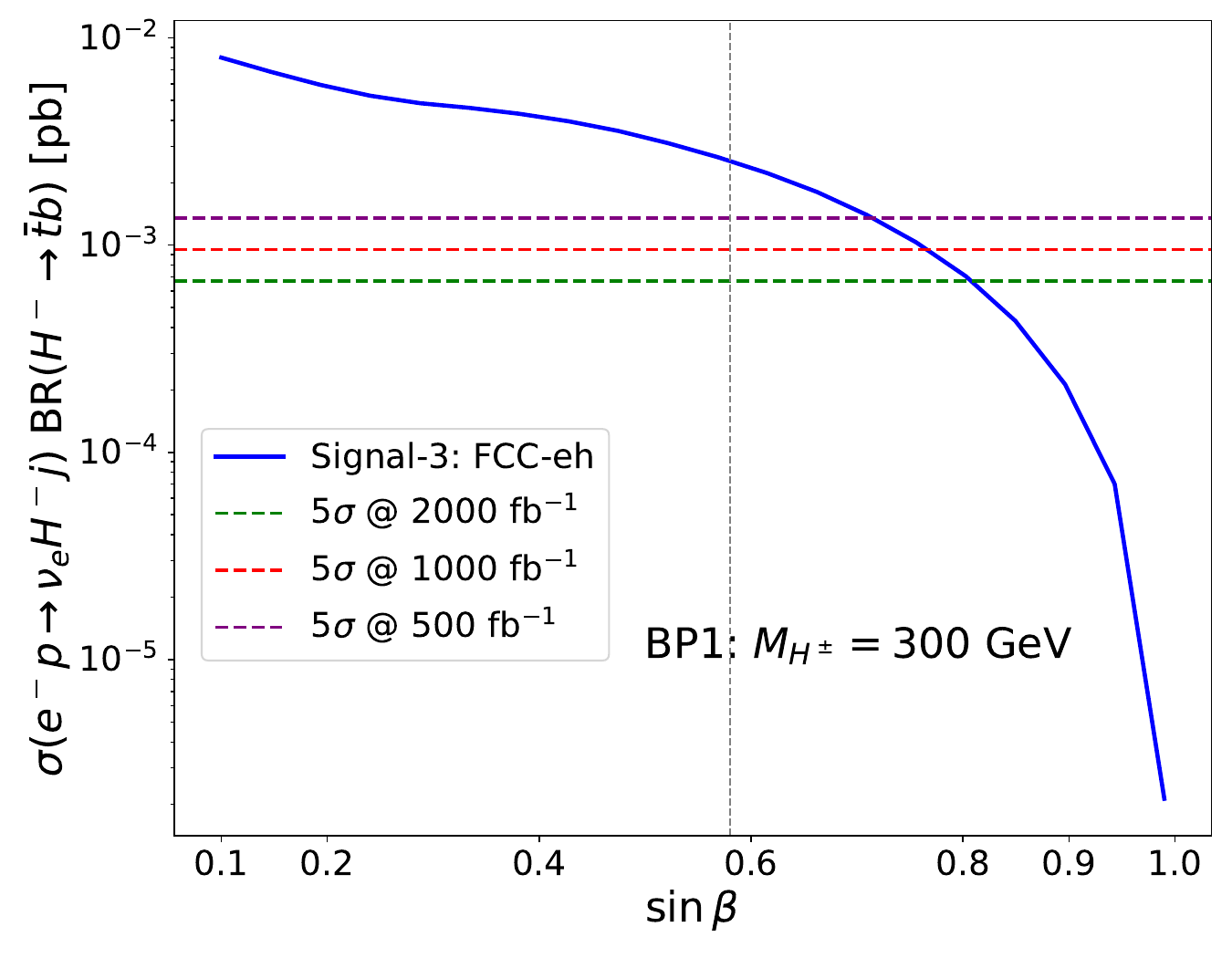}
    \includegraphics[width=0.32\linewidth]{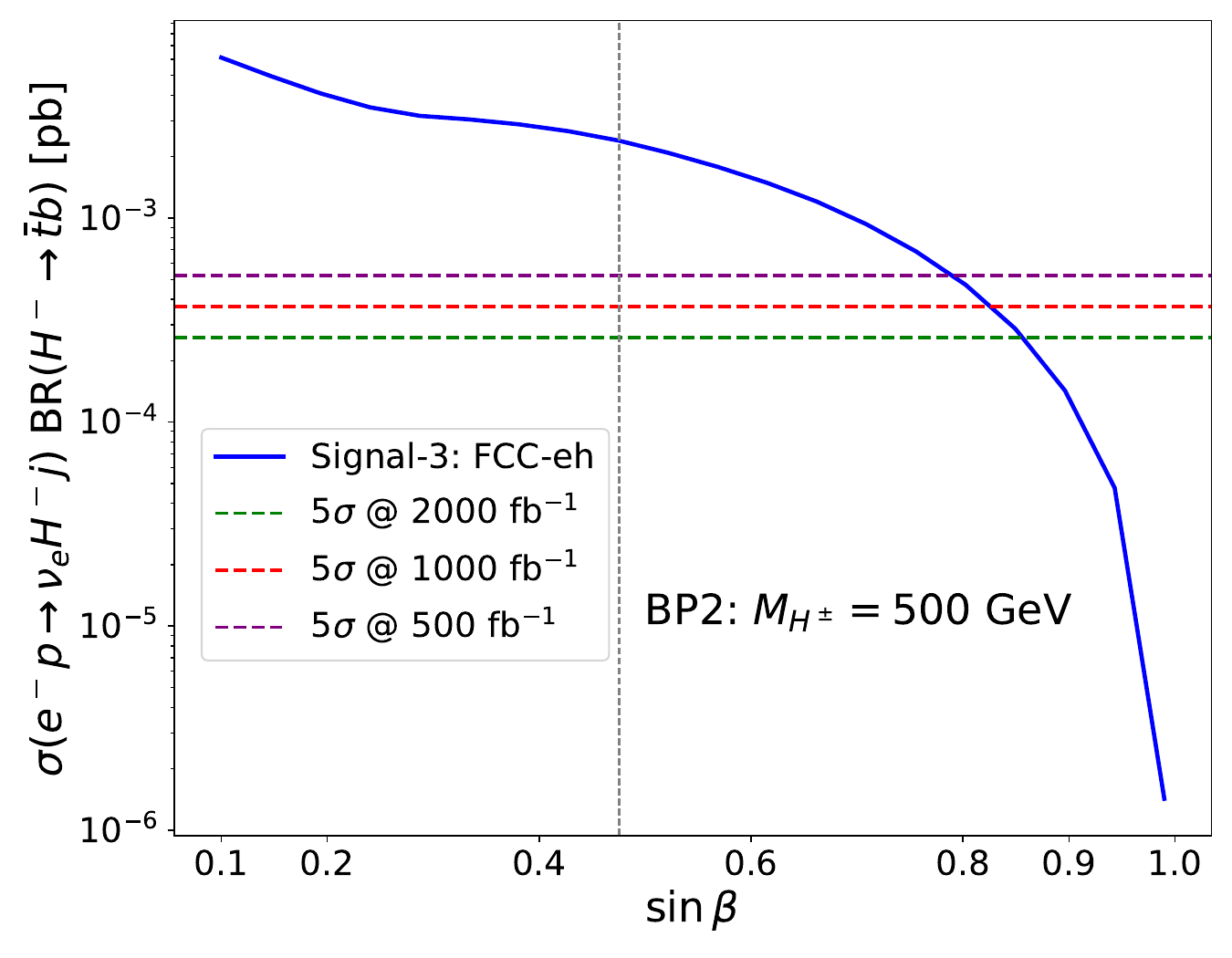}
    \includegraphics[width=0.32\linewidth]{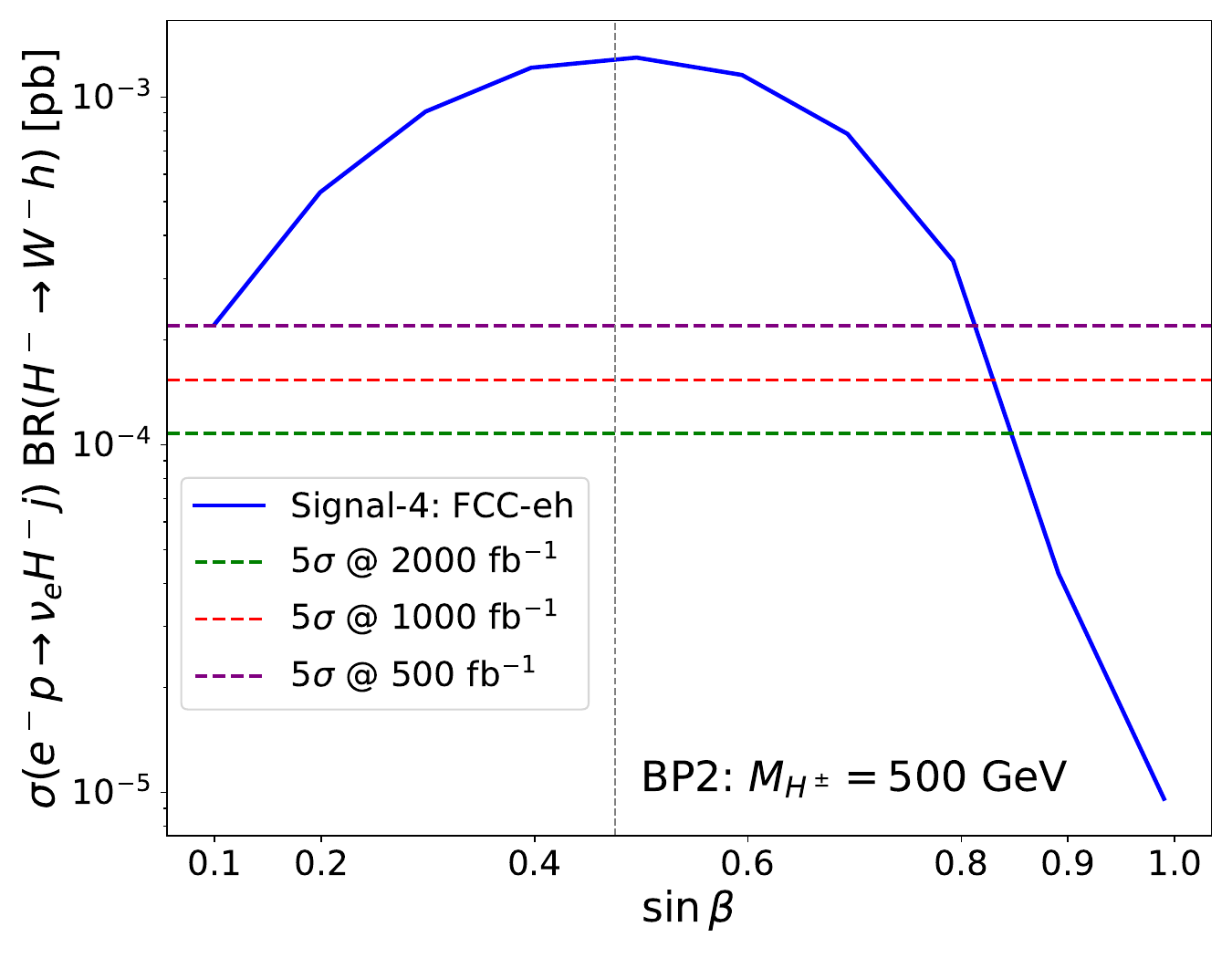}
    \caption{Reach plots for the FCC-eh setup showing the parameter space accessible for all four signal scenarios considered in this study, for different values of $\sin{\beta}$ with $\sin{\alpha} = \tfrac{-1}{\sqrt{2}}$. The region to the left of the gray dashed vertical line is excluded by $b \to s\gamma$ constraints.}
    \label{fig:RP-FCCeh}
\end{figure}

With these considerations, it is natural to focus on the two most relevant signal channels, where the corresponding decay modes consistently provide better sensitivity compared to the remaining ones. Although a more intriguing channel would involve the decay $H^- \to W'Z$, which follows a clear New Physics (NP) $\rightarrow$ NP $\rightarrow$ SM signature chain, its discovery prospects are rather limited. A brief discussion of such channels is provided in Appendix~\autoref{app}, where we demonstrate that they cannot be effectively probed at the future LHeC and FCC-eh setups. Instead, we concentrate on alternative decay modes that offer a realistic possibility of being explored at these colliders. Taking into account the projected luminosities of the proposed LHeC and FCC-eh facilities, we evaluate the discovery potential for the selected benchmark signals. The corresponding reach plots for various benchmark scenarios within this extended gauge model are shown in \autoref{fig:RP-LHeC} for the LHeC setup, and in \autoref{fig:RP-FCCeh} for the FCC-eh setup.

It can be observed that a broad range of $\sin{\beta}$ values allows for the discovery of the charged Higgs in these channels within the integrated luminosity projected for the LHeC (up to $1000~\mathrm{fb}^{-1}$) and the FCC-eh (in the range of $1000$--$2000~\mathrm{fb}^{-1}$). For this analysis, we have also imposed a signal cut efficiency of $\epsilon_s = 0.20$. The Higgs mixing angle is fixed to $\sin{\alpha} = -\tfrac{1}{\sqrt{2}}$, a choice consistent with both theoretical and experimental constraints calculated for this model. At the LHeC setup, a more refined analysis reveals that signal~2, signal~3, and signal~4 exhibit a sufficient range of $\sin{\beta}$ values that provide viable discovery prospects, though their effectiveness varies across the BPs. Specifically, signal~2 shows discovery potential only at BP-2, signal~3 is promising at both BP-1 and BP-2, while signal~4 is viable only at BP-2. In contrast, signal~1 at the LHeC does not yield a sufficient number of signal events after all selection cuts to claim discovery. On the other hand, at the FCC-eh, all four signal channels (signal~1 to signal~4) demonstrate promising discovery potential within the explored parameter space. In particular, signal~1 and signal~3 are viable at both BP-1 and BP-2, whereas signal~2 and signal~4 show significant prospects only at BP-2.

\section{Conclusion}
\label{tab:conclusion}

While current experiments such as the LHC have not yet revealed any clear evidence of new physics beyond the SM, it is essential to explore and test well-motivated BSM scenarios at future collider facilities. Upcoming projects and upgrades, such as the LHeC and FCC-eh, offer unique advantages that complement the LHC program. It is therefore natural to investigate such models in these future avenues as well.

The extended gauge model discussed in this work possesses distinctive signatures that are not typically present in other BSM frameworks. At the high-luminosity LHC, previous studies have shown that the charged Higgs boson in this model can be probed in unique channels such as $H^\pm \to W^{'\pm}Z$, where the decay proceeds via a clear New Physics (NP) $\to$ NP $\to$ SM cascade~\cite{Coleppa:2021wjx}. Motivated by these results, we explored the phenomenology of this model at the LHeC and FCC-eh, where the production mechanisms and kinematic regimes are inherently different. 

We find that at the LHeC and FCC-eh, the exotic channel $H^\pm \to W'Z \to$ SM particles, as discussed in~\cite{Coleppa:2021wjx}, cannot be effectively probed using a realistic cut-flow strategy. The details of this limitation are briefly outlined in Appendix~\autoref{app}. Nevertheless, more conventional decay modes such as $H^- \to \bar{t}b$ and $H^- \to W^-h$ remain highly promising. At the LHeC, operating at $\sqrt{s} = 1.3~\mathrm{TeV}$ with an integrated luminosity of $1000~\mathrm{fb}^{-1}$, we find viable discovery prospects in signal~2,3 and signal~4, corresponding to the final state $2b + j + \ell^- + \cancel{E}_T$. 

Furthermore, at the FCC-eh, with its enhanced center-of-mass energy $\sqrt{s} = 3.5~\mathrm{TeV}$ and projected luminosity in the range of $1000$--$2000~\mathrm{fb}^{-1}$, all four signal channels demonstrate promising discovery prospects within the explored parameter space. In particular, signal~1 and signal~3 are viable at both benchmark points, while signal~2 and signal~4 show significant discovery potential only at BP-2. We conclude by emphasizing that models with an extended gauge and scalar sector remain a fertile ground for rich phenomenology. They offer unique signatures that can be probed across multiple collider environments, including both future facilities and upgrades to current experiments. Such studies provide strong motivation for pursuing new avenues in the search for physics beyond the Standard Model in the near future.

\section*{Acknowledgements}

G.B.K. gratefully acknowledges the initial financial support provided under the Prime Minister's Research Fellowship (Grant No.~PMRF-192002-1802), awarded by the Ministry of Education, Government of India. The author also thanks Dr.~Agnivo Sarkar for valuable discussions and helpful suggestions.

\section*{Data Availability Statement}

There are no publicly available datasets associated with this manuscript. All the data used in this study were generated through Monte Carlo simulations. The list of signal processes and relevant Standard Model background channels, along with the details of the simulation framework and packages employed for event generation, are clearly described in the manuscript. Given that the datasets can be straightforwardly reproduced using the information provided, and considering the large volume of simulated events involved in this analysis, we do not deposit the data separately. Interested readers can regenerate the necessary data following the methodology outlined in the paper.

\bibliography{reference}

\section{Appendix}
\label{app}

\subsection{Phenomenology of the NP $\to$ NP $\to$ SM Cascade Decay: $H^\pm \to W^{'\pm}Z \to W^{\pm}Z\,Z$}

A notable feature of this model is the presence of a non-zero $\xi^{H^{\pm}}_{W^{\mp}Z}$ coupling at tree level, which is absent in models where electroweak symmetry breaking arises solely from scalar singlets or doublets (see Eq.~\ref{Eq:Xirho} and Ref.~\cite{Johansen:1982qm}). In principle, this allows the decay channel $H^{\pm} \to WZ$. However, this mode is suppressed by an additional factor of $x^{2}$, making it phenomenologically difficult to observe at future colliders. In contrast, the coupling between $H^{\pm}$ and $W'Z$ depends only on $\sin\beta$ and is therefore not suppressed. This makes the $H^{\pm} \to W'Z$ decay an attractive new-physics (NP) signature in this framework.  

This distinctive behavior originates from the following interaction term in the effective Lagrangian:
\begin{equation}
\mathcal{L} = \xi\, H^\pm\, W_\mu^\mp\, Z^\mu + \text{h.c.}, 
\end{equation}
where the effective coupling $\xi$ can be written in a general form as
\begin{equation}
\xi^2 = \frac{g^2}{4 m_W^2} 
\left[ \sum_i Y_i^2 \left( 4 T_i (T_i + 1) - Y_i^2 \right) v_i^2 \right] - \frac{1}{\rho^2},
\label{Eq:Xirho}
\end{equation}
with $\rho = \dfrac{m_W^2}{m_Z^2 \cos^2 \theta_W}$. Here, $T_i$ and $Y_i$ denote the isospin and hypercharge quantum numbers of the $i^\text{th}$ Higgs field, while $v_i$ is its vacuum expectation value (vev).  

\vspace{0.3em}
\noindent\textbf{Signal topology:}  
The NP-to-NP-to-SM cascade decay we considered is
\[
e^-p \to \nu_e H^- j,\quad 
H^- \to W'^-Z,\quad 
W'^- \to W^-Z,\quad 
W^- \to jj,\quad 
Z \to b\bar{b}.
\]

This leads to the characteristic final state  
\[
4b + 3j + \cancel{E}_T,
\]  
which we investigate below. Based on this final-state topology and the $e^-p$ collider environment, the relevant SM background processes are summarized in \autoref{tab:SMBG_4b3jMET}. For the signal analysis, we apply baseline object multiplicity cuts, namely $N(b) \geq 4$ and $N(j) \geq 3$. The resulting cutflow and event reduction efficiencies for each benchmark scenario are given in \autoref{tab:npbp1} and \autoref{tab:npbp2}.  

\begin{table}[h]
    \centering
    \begin{tabular}{ !{\vrule}c !{\vrule} c !{\vrule} c !{\vrule}} 
        \toprule
        \textbf{Background Label} & \textbf{Background Process} & \textbf{Final State} \\ 
        \midrule
        BG1 & $e^-p \rightarrow \nu_e\, t\bar{t}\, j$ & $2b + 5j + \cancel{E}_T$ \\ 
        \midrule
        BG2 & $e^-p \rightarrow \nu_e\, b\bar{b}\, b\bar{b}\, j$ & $4b + j + \cancel{E}_T$ \\
        \midrule
        BG3 & $e^-p \rightarrow \nu_e\, Z\, b\bar{b}\, j,\ Z \rightarrow b\bar{b}$ & $4b + j + \cancel{E}_T$ \\
        \midrule
        BG4 & $e^-p \rightarrow \nu_e\, W\, Z\, j,\ W \rightarrow jj,\ Z \rightarrow b\bar{b}$ & $2b + 3j + \cancel{E}_T$ \\
        \midrule
        BG5 & $e^-p \rightarrow \nu_e\, t\, j$ & $b + 2j + \cancel{E}_T$ \\
        \bottomrule
    \end{tabular}
    \caption{Relevant SM background processes for the signal final state \(4b + 3j + \cancel{E}_T\) at the electron-proton collider.}
    \label{tab:SMBG_4b3jMET}
\end{table}

\vspace{0.3em}

\begin{table}[h]
\centering
\begin{minipage}{0.48\textwidth}
    \centering
    \begin{tabular}{!{\vrule}c!{\vrule}c!{\vrule}c!{\vrule}c!{\vrule}c!{\vrule}c!{\vrule}c!{\vrule}}
        \toprule
        \textbf{Cuts} & \textbf{Signal} & \textbf{BG1} & \textbf{BG2} & \textbf{BG3} & \textbf{BG4} & \textbf{BG5} \\
        \midrule
        -- & 100000 & 100000 & 100000 & 100000 & 100000 & 300000 \\
        
        \( N_b \geq 4 \) & 2074.9 & 56.01 & 50.5 & 152.5 & 38.5 & 30.0 \\
        
        \( N_j \geq 3 \) & 831.2 & 31.50 & 2.0 & 5.0 & 4.0 & 0.0  \\
        
        $P_T(j_1)\geq 80$ & 508.7 & 23.0 & 1.0 & 2.0 & 4.0 & 0.0  \\
        
        $M_{b_1b_2j_1j_2}\geq 250$ & 506.6 & 23.0 & 1.0 & 2.0 & 4.0 &  0.0 \\
        
        $H_T\geq 300$ & 503.5 & 23.0 & 1.0 & 2.0 & 4.0 & 0.0 \\
        \bottomrule
    \end{tabular}
    \caption{Cutflow for signal $NP \to NP \to SM$ in BP-1 ($M_{H^\pm}=450~\text{GeV}$, $M_{W'}=350~\text{GeV}$).}
    \label{tab:npbp1}
\end{minipage}
\hfill
\begin{minipage}{0.48\textwidth}
    \centering
    \begin{tabular}{!{\vrule}c!{\vrule}c!{\vrule}c!{\vrule}c!{\vrule}c!{\vrule}c!{\vrule}c!{\vrule}}
        \toprule
        \textbf{Cuts} & \textbf{Signal} & \textbf{BG1} & \textbf{BG2} & \textbf{BG3} & \textbf{BG4} & \textbf{BG5} \\
        \midrule
        -- & 100000 & 100000 & 100000 & 100000 & 100000 & 300000 \\
        
        \( N_b \geq 4 \) & 2367.9  & 56.01 & 50.5  & 152.5 & 38.5 & 30.0 \\
        
        \( N_j \geq 3 \) & 1127.7 & 31.50 & 2.0 & 5.0 & 4.0 & 0.0 \\
        
        $P_T(j_1)\geq 100$ & 514.7 & 18.0 & 1.0 & 1.0 & 3.5 & 0.0 \\
        
        $M_{b_1b_2j_1j_2}\geq 300$ & 396.1 & 15.5 & 1.0 & 0.5 & 1.0 &  0.0 \\
        
        $H_T\geq 350$ & 396.1 & 15.5 & 1.0 & 0.5 & 1.0 & 0.0 \\
        \bottomrule
    \end{tabular}
    \caption{Cutflow for signal $NP \to NP \to SM$ in BP-2 ($M_{H^\pm}=550~\text{GeV}$, $M_{W'}=350~\text{GeV}$).}
    \label{tab:npbp2}
\end{minipage}
\end{table}

\begin{figure}[H]
    \centering
    \includegraphics[width=0.45\linewidth]{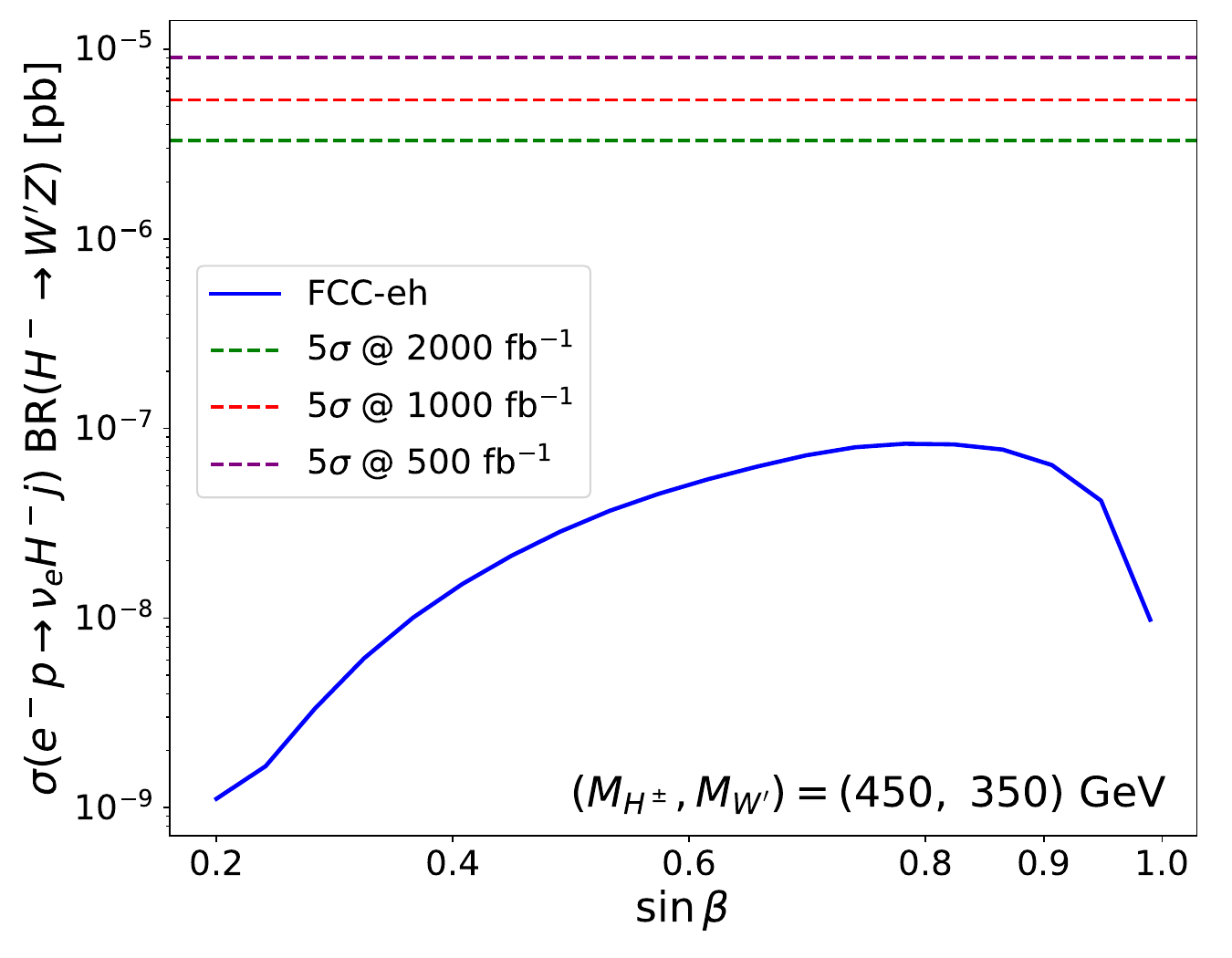}
    \includegraphics[width=0.45\linewidth]{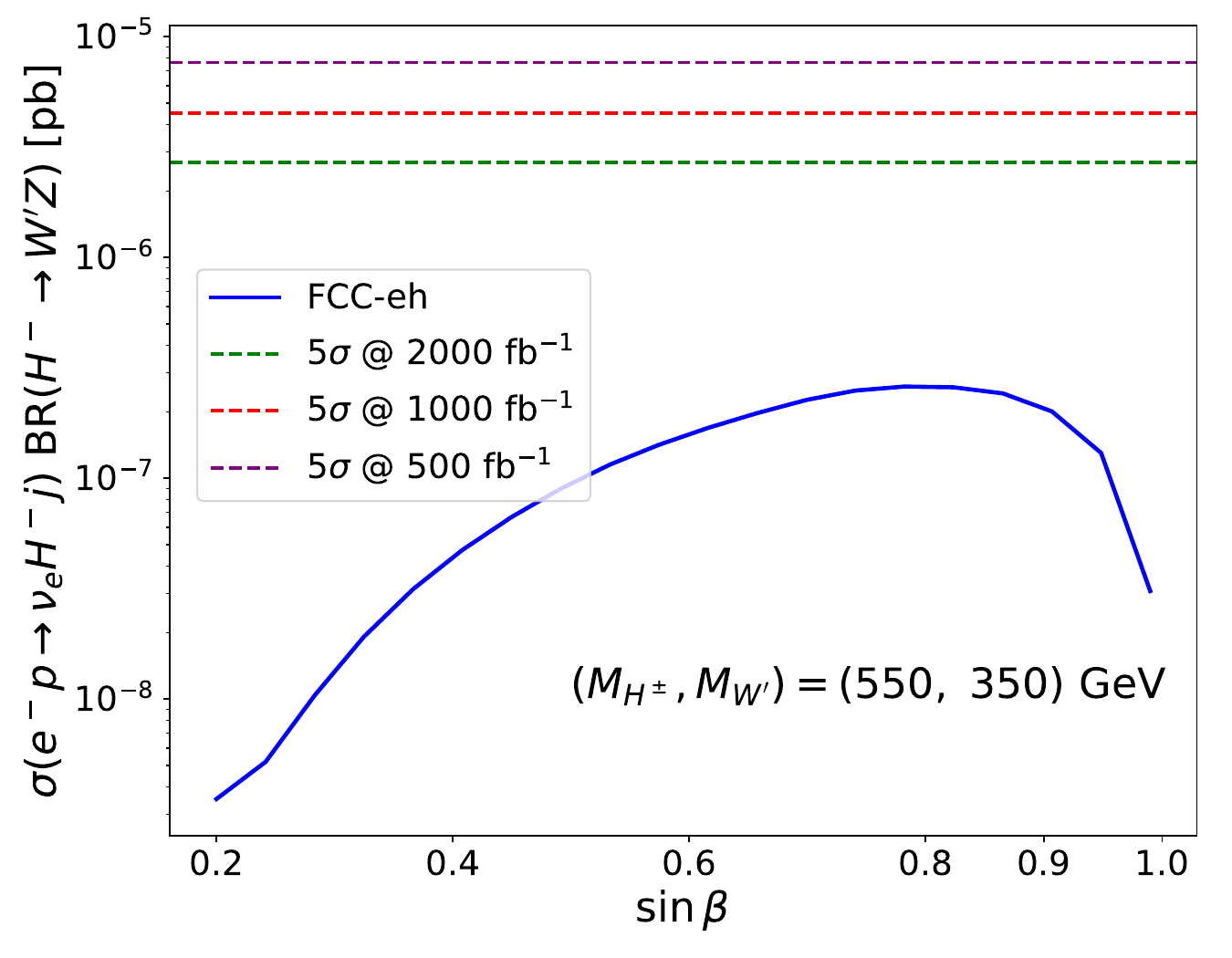}
    \caption{Estimated signal reach for the NP$\to$NP$\to$SM decay mode at the FCC-eh for the two benchmark points.}
    \label{fig:NPNP}
\end{figure}

\vspace{0.3em}
Using these optimized cutflows, we estimate the required signal cross sections for both discovery and exclusion. For the LHeC setup, even after accounting for the expected integrated luminosity, the resulting event yield remains below 5 events, making this channel inaccessible. For FCC-eh, despite the higher center-of-mass energy and luminosity, the reach plots indicate that this NP$\to$NP$\to$SM cascade decay mode remains challenging to probe. The corresponding reach plots are shown in \autoref{fig:NPNP}.  

\end{document}